\documentclass[nofootinbib,nobibnotes,groupaddress,preprintnumbers,aps,pre,twocolumn]{revtex4}
\usepackage{tipa}
\usepackage{color}
\usepackage{bm,epsfig,mathrsfs,amsmath,amssymb,graphicx,subfigure,overpic}
\usepackage{graphicx}
\usepackage{dcolumn}
\usepackage{bm}
\usepackage{chngcntr}
\usepackage{float} 
\usepackage{booktabs}%
\newcommand{\PreserveBackslash}[1]{\let\temp=\\#1\let\\=\temp}
\newcolumntype{C}[1]{>{\PreserveBackslash\centering}p{#1}}
\newcolumntype{R}[1]{>{\PreserveBackslash\raggedleft}p{#1}}
\newcolumntype{L}[1]{>{\PreserveBackslash\raggedright}p{#1}}
\usepackage[colorlinks,citecolor=blue,linkcolor=red]{hyperref}
\usepackage[toc,title,titletoc,header]{appendix}

\date{\today}
\begin{document}
\title{Quantum Otto engine with quantum correlations}
\author{Yang Xiao$^{1}$}
\author{Dehua Liu$^1$}
\author{Jizhou He$^{1}$}
\author{Yongli Ma$^2$}\email{ylma@fudan.edu.cn}
\author{Zhaoqi Wu$^3$}
\author{Jianhui Wang$^{1,2}$}\email{wangjianhui@ncu.edu.cn}
\affiliation{ $^1\,$ Department of Physics, Nanchang University,
Nanchang 330031, China\\ $^2\,$   State Key Laboratory of Surface
Physics and Department of Physics, Fudan University, Shanghai
200433, China\\ $^3\,$  School of Mathematics and Computer Science, Nanchang University, Nanchang
330031, China}

\begin{abstract}
We theoretically propose and investigate a quantum Otto engine that works with a single-mode radiation field inside an optical cavity and is driven alternately by a hot reservoir and a cold reservoir. The hot reservoir is realized using a beam composed of thermally entangled pairs of two-level atoms that interacts resonantly with the cavity, and the cold reservoir is composed of a collection of noninteracting boson modes. In terms of the quantum discord of the pair of atoms, we derive
analytical expressions for the performance parameters (i.e., power and efficiency) and the stability measure (the coefficient of variation of power). We show that nonclassical correlations boost the quantum engine’s performance and efficiency, and they may even change the operation mode {at specified values of the two bath temperatures}. We also demonstrate that the nonclassical correlations improve the stability of the machine by reducing the coefficient of variation of power that satisfies the generalized thermodynamic uncertainty relation. Finally, we find that these results can be transferred to another quantum Otto engine model, in which the optical cavity is coupled alternately to a hot thermal bosonic bath and to a beam composed of pairs of the two correlated atoms that plays the role of a cold reservoir. 

\end{abstract}
\maketitle
\date{\today}
\section{Introduction}

The presence of nonclassical correlations is one of the most intriguing signatures of the nonclassicality of a quantum state.
These correlations have been considered to be a type of resource in certain quantum tasks, including quantum computation and quantum information processing \cite{Hor09,Ani08,And19,Per15,Dat07,Lan08, Son19,Chen13}. Quantum discord, which is generally used to quantify nonclassical correlations, is more general than quantum entanglement \cite{Modi12, Son19} and is defined as the difference between the quantum mutual
information and the classical correlations in a system \cite{Zurek01,Rau10, Son19}. Although quantum discord is not always greater than quantum entanglement, it may be nonzero even for separable states in the absence of entanglement \cite{Ved01,Zurek01,Ani08,Lutz09}. The quantum discord represents a good figure of merit for
quantum resource characterization, not only in quantum information theory, but also in the quantum thermodynamics field \cite{Hew18, Bar17, Alt14,wang22} with a focus on quantum thermal machines \cite{ Bar17, Alt14}.

{One topic of major interest in quantum thermodynamics is the finite-time performance of nanoscale heat engines, in which quantum effects are apparently held in the working substance \cite{Bar17, Alt14, Huang13, Par13, Hew18, NJP1618,chen18} and may even govern the engine’s reservoirs \cite{Aba14,Kla17,RJ19}. In other words,
 in the quantum realm, both the working substance and the heat reservoirs can be composed of finite-sized systems that can be prepared in quantum states without classical analogues. Finite-time quantum heat engines that work with interacting working systems have been observed to have performance characteristics that are quite different to those with working systems that consist of ideal simple systems \cite{Alt14,chen18}. While the quantum correlations that are inherent in the working medium may boost the performance of some quantum heat engines \cite{NJP1618},  the nonclassical correlations that exist in the working substance may degrade the machine performance of other engines \cite{Alt14,Hong20}.} 

When they play the roles of the reservoirs, these finite-dimensional systems can lead to a nonthermal scenario in which the heat engines outperform their classical analogs \cite{Aba14,Kla17,RJ19}. Additionally, the fluctuations in these quantum heat engines cannot be neglected because both the work and the heat are stochastic and fluctuate \cite{Sei12,Sek10,Jiang15,Lutz20,Esp14,Liu20,Tu22,Hol121}. Quantum heat engines that operate with out-of-equilibrium reservoirs, which may be squeezed \cite{Wang19,Aba14,Kla17,Sin20,You18,Manz18,Nie16,Assis20,Manz16, Pap22}, quantum coherent \cite{Scu11,Quan06,Scu03,Kor16,Guff19,Rod19,Str17}, quantum measurement induced \cite{Tal17,Ding18,Su21,Buf19,Her17,Jor18}, and
nonclassically correlated \cite{Ber17,And19,Per15}, were investigated theoretically and demonstrated experimentally in either finite-time or quasi-static operating modes. These engines, which outperform their classical counterparts, may be models based on different cycles, e.g., the Carnot cycle \cite{Scu03, Quan06}, the Otto cycle \cite{Aba14, Wang19, Kla17}, and the Stirling cycle \cite{Pap22}. A photo-Carnot engine that was introduced by Scully and co-workers \cite{Scu03} was extended by including quantum coherence to explore
the engine’s thermodynamic efficiency beyond the Carnot limit \cite{Lutz09}. It is well known that, with the exception of the case of the reversible Carnot engine working between two thermal reservoirs \cite{Car24,Cal85}, the performance of heat engines is always model dependent, irrespective of the use of thermal or nonthermal reservoirs.

  In this paper, with the aim of using quantum correlations in thermodynamic applications, we propose a quantum Otto engine, which has not been studied to date. The proposed engine works with a single-mode optical cavity and is driven using an out-of-equilibrium reservoir with quantum correlations. 
  In addition to two thermodynamic adiabatic processes, the proposed engine consists of hot and cold isochoric strokes, where the optical cavity is weakly coupled to a thermal reservoir during the cold isochore and interacts resonantly with one of the pair of correlated atoms along the hot isochore. {
      It may be more difficult to prepare the entangled pair of atoms with only one being coupled to the optical cavity that plays the role of the reservoir than to realize the entangled modes under fully noninteracting reservoir conditions. However, it is important to consider such a physical model because it provides a platform for understanding of nanoengines in which the quantum effects of the reservoirs are involved.}

  We examine the finite-time machine performance and the fluctuations by varying the quantum discord. We show that {although the machine will operate as a heat engine at certain bath temperatures with any value of quantum discord}, a modulation of the quantum discord may change the operation mode, e.g., a thermal device that is expected to operate as a refrigerator in the absence of quantum discord may operate as a heat engine when the nonclassical correlations are present. For the heat engine, we find that these nonclassical correlations (i) increase both work extraction and thermodynamic efficiency but reduce the relative power fluctuations, with the latter satisfying the generalized thermodynamic uncertainty relation \cite{Tim19}, and (ii) enable the finite-time engine to operate at efficiencies beyond the Carnot limit. Finally, we demonstrate that these results are applicable to an alternative quantum Otto cycle, in which the hot reservoir is composed of quantized boson modes and the cold reservoir consists of the pairs of two-level atoms that pass sequentially through the optical cavity.

{The remainder of this paper is organized as follows. We briefly review quantum discord and quantum entanglement and clarify the difference between these two quantities in Sec. \ref{QE}, where we also explain why the quantum discord rather than the concurrence is used to describe the nonclassical correlations. In Sec. \ref{QO}, the performance of the quantum Otto engines is investigated. We describe the Otto cycle model in Sec. \ref{QOf} and subsequently discuss the machine performance and the power fluctuations in Sec. \ref{QOm}, where the effects induced by quantum discord are also explored. 
   Finally, we discuss the results and draw conclusions in Sec. \ref{Co}. This paper also includes four appendices: Appendix \ref{aped}, derivation of the nonadiabatic factor; Appendix \ref{apet}, discussions of the time durations along the two isochoric
processes; Appendix \ref{apea}, numerical analysis of an alternative Otto engine in which the optical
cavity is coupled to the pair of interacting atoms
along the cold isochore; and Appendix \ref{apeq}, results with respect to the quantum Otto engines using the two-atom system with nonclassical correlations as the working substance.}

\section{Quantum entanglement and discord}\label{QE}
Consider the Hamiltonian of two identical spin-$1/2$ atoms with frequency $\omega$ via an $XY$ Heisenberg interaction ($\hbar\equiv1$) \cite{zheng00}:
\begin{equation}\label{ham}
H^{sp}=\frac{\omega}{2}(\sigma_1^z+\sigma_2^z)+\xi(\sigma_1^+\sigma_2^-+\sigma_1^-\sigma_2^+),
\end{equation}
where $\xi$ is the controllable strength of the interaction. The operator $\sigma_j^+=(\sigma_j^-)^{\dag}=|e\rangle_{jj}\langle g|=\frac{1}{2}(\sigma^{x}_j+i\sigma^{y}_{j}),
  \sigma_j^z=|e\rangle_{jj}\langle e|-|g\rangle_{jj}\langle g|$,
 where $|g\rangle_j$, $|e\rangle_j$, and $\sigma^{\alpha}_j$ ($\alpha=x,y,z$) are the ground state, the excited state, and the Pauli matrices for atom $j$ (with $j=1,2$), respectively.
The eigenvalues $E_i$ and the corresponding eigenvectors $|\Psi_i\rangle$ of the Hamiltonian in Eq. (\ref{ham}) can be calculated as
\begin{eqnarray}\label{eigen}
E_1&=&-E_4=\omega, |\Psi_1\rangle=|ee\rangle, |\Psi_4\rangle=|gg\rangle,\nonumber\\
E_2&=&-E_3=\xi,  |\Psi_2\rangle=\frac{\sqrt{2}}{2}(|ge\rangle+|eg\rangle),\\
|\Psi_3\rangle&=&\frac{\sqrt{2}}{2}(-|ge\rangle+|eg\rangle)\nonumber.
\end{eqnarray}

When the two-atom system is at thermal equilibrium with an inverse temperature $\beta$, where $\beta=1/(k_BT)$, the density operator can be determined using $\rho_{12}=e^{-\beta H^{sp}}/Z^{sp}$, where the canonical partition function for the spin system is $Z^{sp}=\mathrm{Tr}(e^{-\beta H^{sp}})$. It then follows, when using Eqs. (\ref{ham}) and (\ref{eigen}), that the thermal state of the system can be described by
\begin{eqnarray}\label{rho12}
\rho_{12}&=&\frac{1}{Z^{sp}}(e^{-\beta \omega} |\Psi_1\rangle\langle \Psi_1|+e^{-\beta \xi} |\Psi_2\rangle\langle \Psi_2|\nonumber\\
&+&e^{\beta \xi} |\Psi_3\rangle\langle \Psi_3|+e^{\beta \omega} |\Psi_4\rangle\langle \Psi_4|),
\end{eqnarray}
 where the partition function $Z^{sp}=2[\mathrm{cosh}(\beta\omega)+\mathrm{cosh}(\beta\xi)]$.
 To describe the entanglement between the two atoms, the concurrence \cite{Woo97,Woot97} is introduced. The concurrence can be evaluated using Eq. (\ref{rho12}) when it is written in a closed form:
\begin{eqnarray}\label{ent}
\mathcal{C}(\rho_{12})=\mathrm{max}\{0,\frac{\mathrm{sinh}(\beta\xi)-1}{\mathrm{cosh}(\beta\omega)+\mathrm{sinh}(\beta\xi)}\}.
\end{eqnarray}
When $\beta\xi\leq \mathrm{arcsinh}(1)$, the concurrence is zero and the mixture is separable because of the vanishing entanglement, but when $\beta\xi>\mathrm{arcsinh}(1)\approx0.88$, the concurrence is positive and the entanglement between the two atoms is non-negligible. This implies that the thermal state (\ref{rho12}) would be entangled in the low-temperature and/or strong-coupling cases.

The quantum discord, which is defined as the difference between the quantum mutual information $\mathcal{I}(\rho_{12})$ and the classical correlations $\mathcal{J}(\rho_{12})$ \cite{Zurek01,Rau10}, takes the form of
\begin{equation}\label{QD}
\mathcal{Q}(\rho_{12}):=\mathcal{I}(\rho_{12})-\mathcal{J}(\rho_{12}).
\end{equation}
The quantum mutual information is given by $\mathcal{I}(\rho_{12})=S(\rho_{1})+S(\rho_{2})-S(\rho_{12})$, where $S(\rho)=-\mathrm{Tr}[\rho\mathrm{log_2\rho}]$ is the von Neumann entropy and $\rho_{1(2)}=\mathrm{Tr}_{2(1)}(\rho_{12})$ is the reduced density matrix of atom $1(2)$. To describe the classical correlations $\mathcal{J}(\rho_{12})$, the measurement basis $\{B_k\}$ is introduced to describe the von Neumann measurement for atom $2$ only. The conditional density operator $\rho_{1}^{k}$ associated with measurement result $k$ is then given by $\rho_{1}^{k}=(I\otimes B_k)\rho_{12}(I\otimes B_k)/p_k$, where the probability $p_k=\mathrm{Tr}[(I\otimes B_k)\rho_{12}(I\otimes B_k)]$. The quantum conditional entropy with respect to this measurement basis is $S(\rho_{1}|\{B_k\})=\sum_{k}p_{k}S(\rho_{1}^{k})$, and the classical correlations are then determined by $\mathcal{J}({\rho_{12}})=\mathop{\mathrm{sup}}\limits_{\{B_k\}}[S(\rho_1)-S(\rho_1|\{B_k\})]$ \cite{vedral03,Luo08,Luo07}. When the parametrized basis $\{B_k\}=\{\mathrm{cos} \theta|g\rangle-\mathrm{sin} \theta|e\rangle,-\mathrm{sin} \theta|g\rangle-\mathrm{cos} \theta|e\rangle\}$ is used, it then follows that the minimum of the discord is reached at $\theta=\pi/4$ \cite{Lutz09}, which reads as follows: 
\begin{eqnarray}\label{Qd}
\mathcal{Q}(\rho_{12})&=&-\frac{1}{\mathrm{ln}(2)}\{2 (\beta\xi)\rho_{nd}+\rho_d\mathrm{ln}[{Z^{sp}}^2(\rho_{g}+\rho_{d})(\rho_{e}+\rho_{d})]\nonumber\\
&+&\mathop{\sum}\limits_{\alpha=g,e}\rho_{\alpha}\mathrm{ln}(\frac{\rho_{\alpha}+\rho_{d}}{\rho_{\alpha}})+\mathop{\sum}\limits_{\epsilon=\pm}\Phi_{\epsilon}\mathrm{ln}\Phi_{\epsilon}
\},
\end{eqnarray}
where $\rho_{g}=\langle gg|\rho_{12}|gg\rangle=\mathrm{exp}(\beta\omega)/Z^{sp}$, $\rho_{e}=\langle ee|\rho_{12}|ee\rangle=\mathrm{exp}(-\beta\omega)/Z^{sp}$,
$\rho_d=\langle eg|\rho_{12}|eg\rangle=\mathrm{cosh}(\beta \xi)/Z^{sp}$,
$\rho_{nd}=\langle eg|\rho_{12}|ge\rangle=\langle ge|\rho_{12}|eg\rangle=-\mathrm{sinh}(\beta \xi)/Z^{sp}$, and $\Phi_{\epsilon}=(1+\epsilon\sqrt{(\rho_e-\rho_{g})^2+4\rho_{nd}^2})/2$. In the high temperature and/or weakly coupling limit, where
  $\beta\xi\ll1$, Eq. (\ref{Qd}) can be simplified to give $\mathcal{Q}(\rho_{12})\simeq(\beta\xi)^2/(8\mathrm{ln}2)$.
\begin{figure}
\centering
\begin{overpic}[width=7cm]{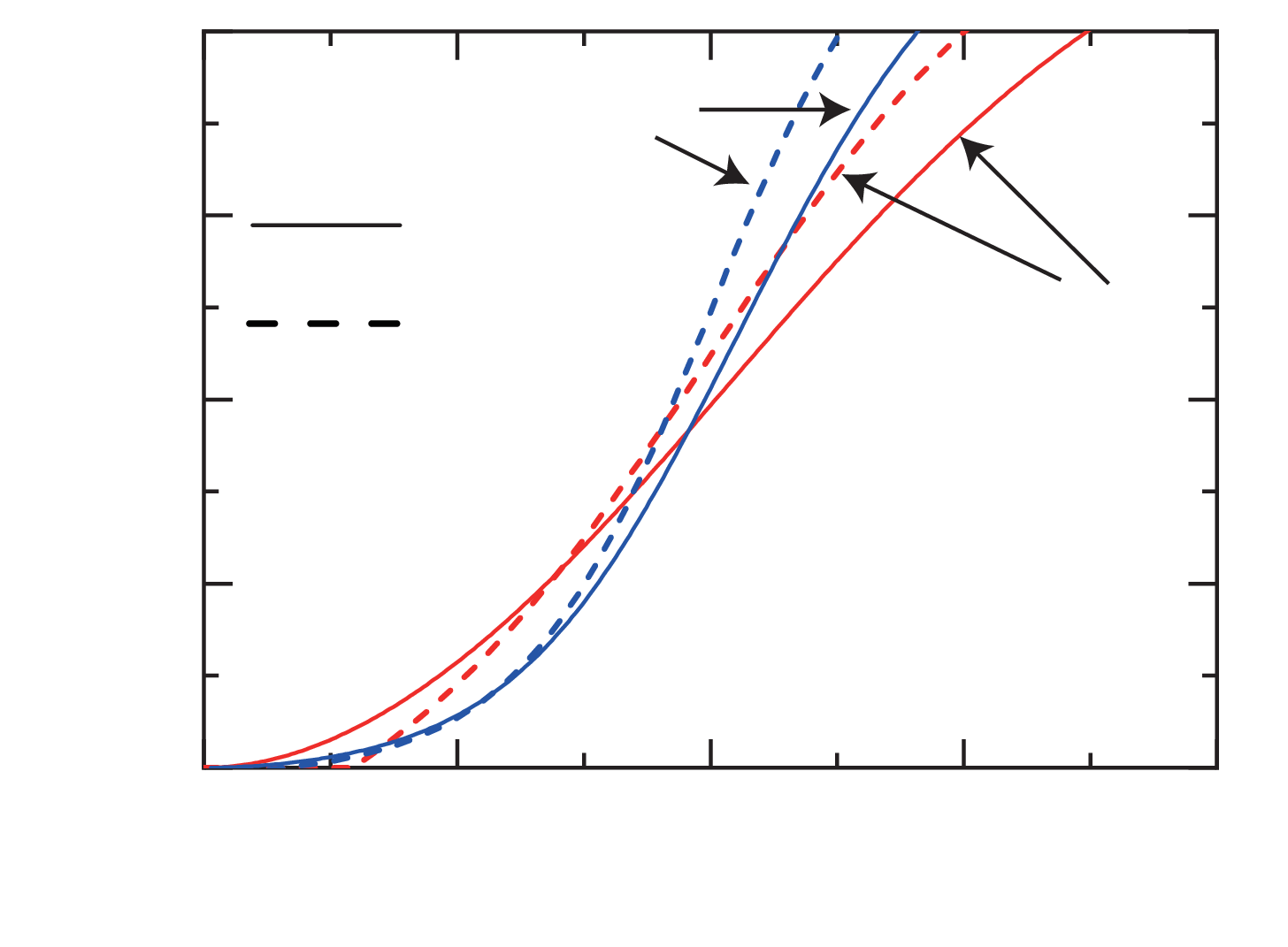}
    \put(14.3,73){(a)}
    \put(7,68){\scalebox{1.3}{0.8}}
    \put(7,55){\scalebox{1.3}{0.6}}
    \put(7,40.5){\scalebox{1.3}{0.4}}
    \put(7,26.5){\scalebox{1.3}{0.2}}
    \put(7,13){\scalebox{1.3}{0.0}}
    \put(14,8){\scalebox{1.3}{0}}
    \put(34,8){\scalebox{1.3}{3}}
    \put(54,8){\scalebox{1.3}{6}}
    \put(73.5,8){\scalebox{1.3}{9}}
    \put(91,8){\scalebox{1.3}{12}}
    \put(33,63.5){\scalebox{1.3}{$\beta=0.9$}}
    \put(73,47){\scalebox{1.3}{$\beta=0.5$}}
    \put(32,55){\scalebox{1.1}{$\mathcal{Q}(\rho_{12})$}}
     \put(33,47){\scalebox{1.1}{$\mathcal{C}(\rho_{12})$}}
     \put(54,3){\scalebox{1.3}{$\xi$}}
\end{overpic}
\begin{overpic}[width=7cm]{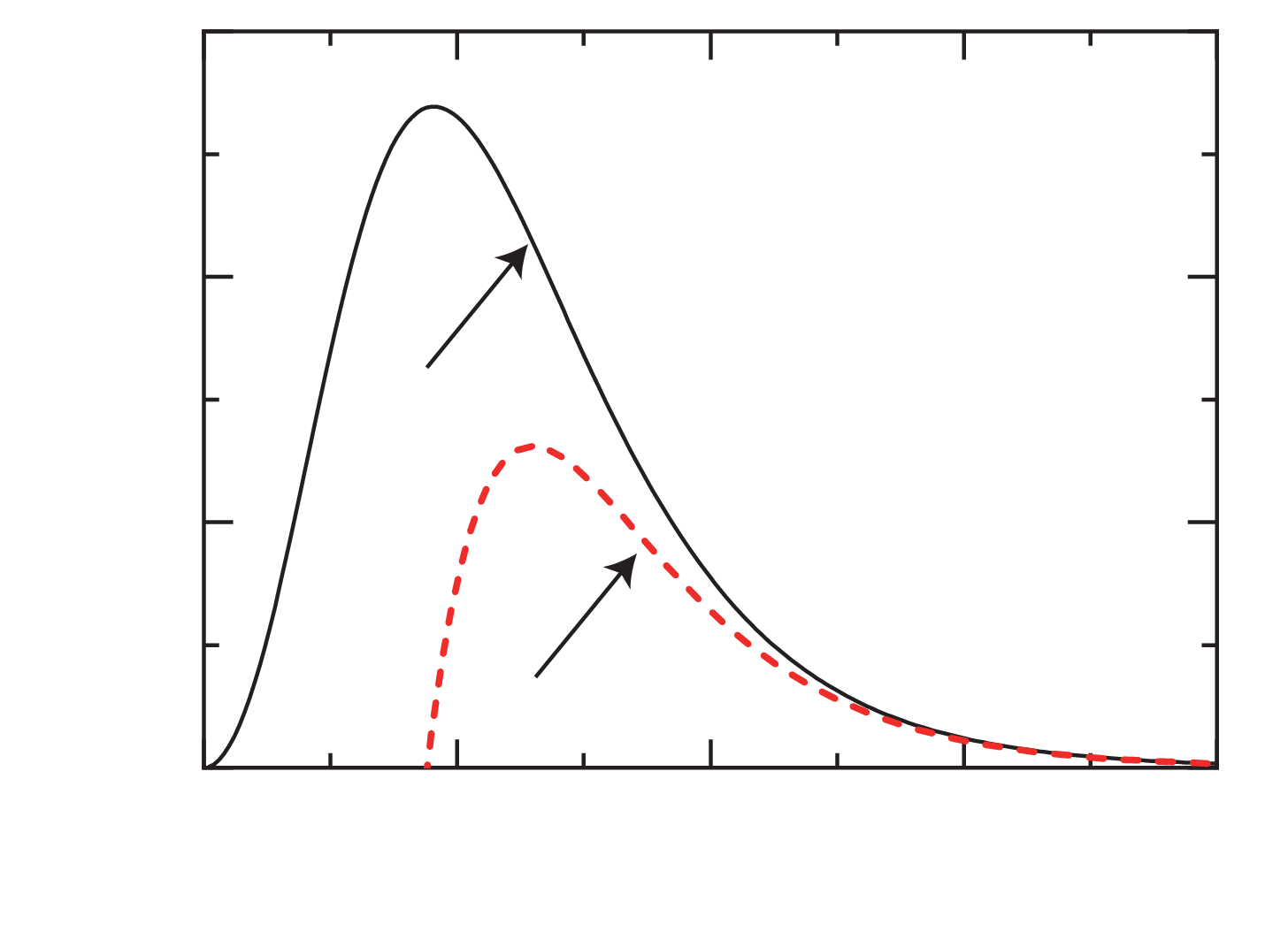}
    \put(14.3,73){(b)}
    \put(4,68){\scalebox{1.3}{0.06}}
    \put(4,50){\scalebox{1.3}{0.04}}
    \put(4,30.5){\scalebox{1.3}{0.02}}
    \put(4,13){\scalebox{1.3}{0.00}}
    \put(14,8){\scalebox{1.3}{0.0}}
    \put(31.5,8){\scalebox{1.3}{0.5}}
    \put(51,8){\scalebox{1.3}{1.0}}
    \put(71,8){\scalebox{1.3}{1.5}}
    \put(91,8){\scalebox{1.3}{2.0}}
    \put(26,41){\scalebox{1.1}{$\mathcal{Q}(\rho_{12})$}}
     \put(35.5,17){\scalebox{1.1}{$\mathcal{C}(\rho_{12})$}}
     \put(54,3){\scalebox{1.3}{$\beta$}}
\end{overpic}
\caption{Quantum discord $\mathcal{Q}(\rho_{12})$ and concurrence $C(\rho_{12})$ as functions of (a) the interaction strength $\xi$ for $\beta=0.5$ and $0.9$, and (b) the inverse temperature $\beta$ for $\xi=2$. The frequency is $\omega=6.$}
    \label{QC}
\end{figure}
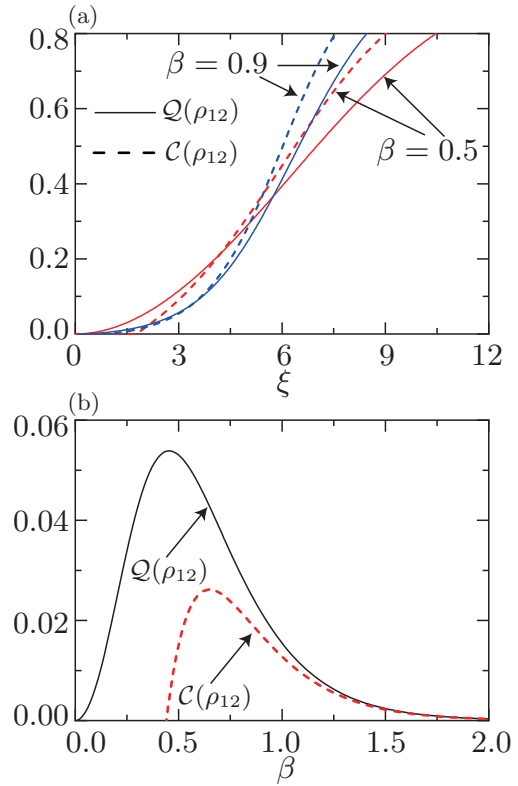

{In
Fig. \ref{QC}(a) we plot the concurrence $\mathcal{C}(\rho_{12})$ and the quantum discord $\mathcal{Q}(\rho_{12})$, which are calculated using Eqs. (\ref{ent}) and (\ref{Qd}), respectively, as functions of
the interaction strength $\xi$ for both $\beta=0.5$ and $\beta=0.9$.} Both the quantum discord and the concurrence, {which depend on the inverse temperature $\beta$}, decrease monotonically as the interaction strength decreases, and they both vanish in the absence of interparticle interactions, as expected. {As the inverse temperature increases, these two quantities begin at zero before rising to their respective maximum values and then decrease again before vanishing at large values of the inverse temperature, as shown in Fig. \ref{QC}(b)}. {Figure \ref{QC}(b) shows that when $\beta\le0.441$, the concurrence is zero, but the quantum discord only vanishes when $\beta\rightarrow 0$, showing that the quantum discord holds in the region in which the concurrence is not observed}. {This means that the quantum discord may be nonzero even when the nonentangled states are separable. Therefore, we will use the quantum discord $\mathcal{Q}(\rho_{12})$ here, rather than the concurrence $\mathcal{C}(\rho_{12})$, to capture the information about the correlations of the two-atom system}.

\section{Quantum Otto engine}\label{QO}
\subsection{Four consecutive strokes in a machine cycle}\label{QOf}
\begin{figure}
\begin{overpic}[width=1.5cm]{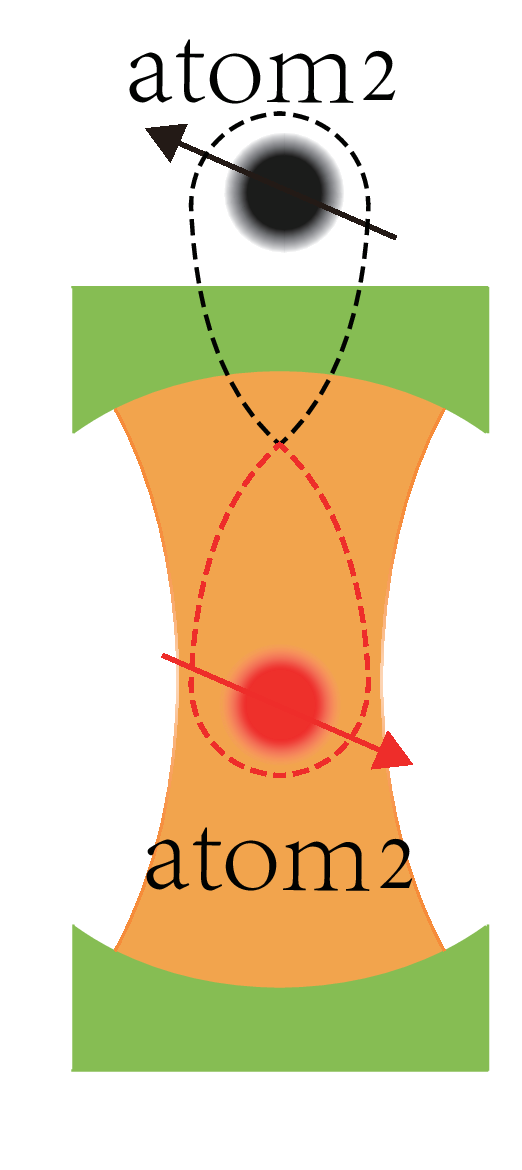}
    \put(-10,90){(a)}
\end{overpic}
\begin{overpic}[width=6.5cm]{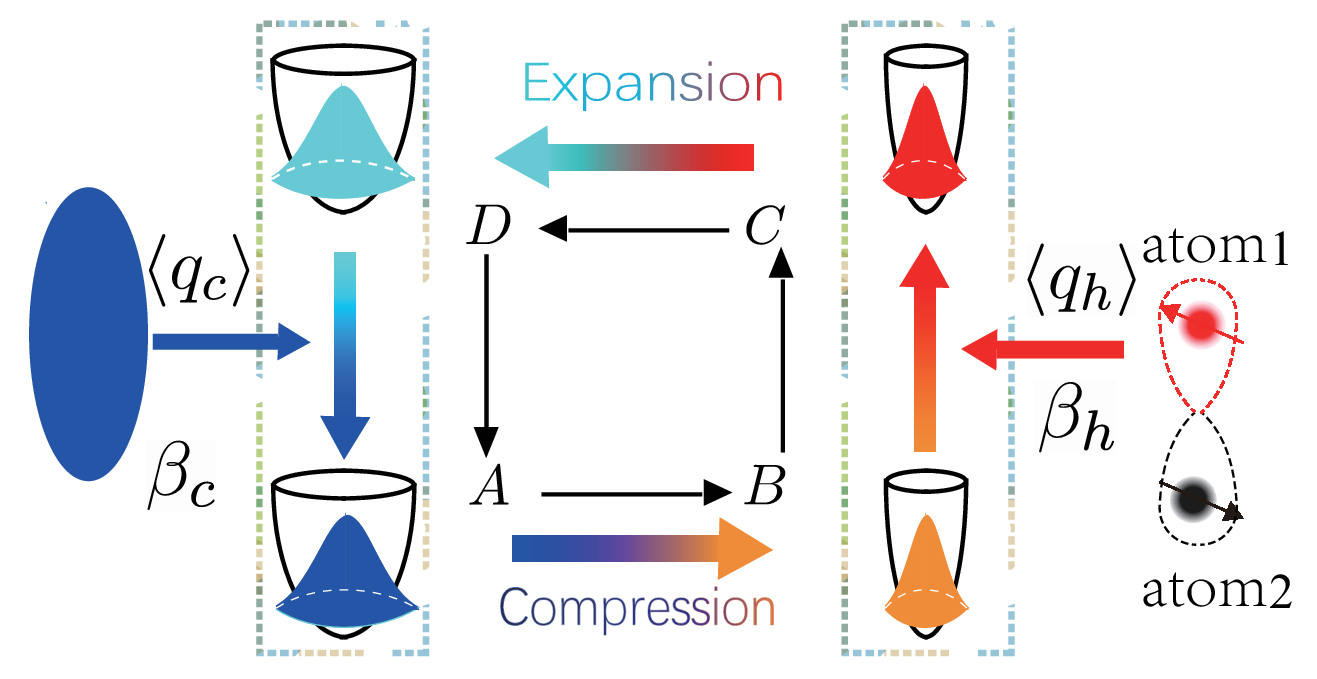}
    \put(8,46.5){(b)}
\end{overpic}
\caption{ (a) Sketch of a single-mode optical cavity coupled to one of two correlated atoms. (b) Illustration of a finite-time quantum Otto engine. During the first stroke $A\rightarrow B$, the working substance at the thermal state undergoes a unitary
compression within the time period $\tau_{ch}$, and its Hamiltonian changes from $H(\omega_{c})$ to $H(\omega_{h})$.
During the second stroke $B\rightarrow C$, the optical cavity for the fixed frequency ($\omega=\omega_h$) is coupled to one of the two correlated atoms with inverse temperature $\beta_h$, and this atom reaches the stationary state at the end of this stroke. The coupling is realized by sending this atom to pass through the optical cavity. 
 The third stroke $C\rightarrow D$ is a unitary expansion in which the Hamiltonian evolves from $H(\omega_{h})$ to $H(\omega_{c})$ during
 the time $\tau_{hc}$. During the cooling stroke $D\rightarrow A$, the working substance with constant frequency $\omega_c$ is in contact with the cold thermal bath of inverse temperature $\beta_{{c}}$ and relaxes to the thermal state by
closing the cycle. The average heat values absorbed by the working system along the hot and cold isochoric strokes are denoted by $\langle q_h\rangle$ and $\langle q_c\rangle$, respectively. We assume that the time spent on the isochoric stroke is much shorter than that spent on the unitary stroke \cite{Aba14} and the time to complete a cycle is given by $\tau_{cyc}=\tau_{ch}+\tau_{hc}$.}
\label{model}
\end{figure}

We consider a quantum engine cycle that uses a single mode of the quantized radiation within an optical cavity as its working substance and operates between hot and cold heat reservoirs. During the cold isochoric process, the thermal bath is composed of an infinite collection of noninteracting bosonic modes, but during the hot isochoric stroke, the optical cavity interacts resonantly with a beam composed of thermally entangled pairs of two-level atoms [see Fig. \ref{model}(a)] that play the role of the hot nonthermal reservoir. 
The interaction of the cavity with these atoms is realized by sending only one of a pair of atoms to pass through the cavity, which means that only one of the atoms in the pair interacts
with the radiation field \cite{Lutz09}. The engine model under consideration as a quantum version of the Otto cycle is drawn schematically in Fig. \ref{model}(b). This model consists of two unitary
strokes denoted by $A\rightarrow B$ and $C\rightarrow D$, where the system is isolated from the two heat reservoirs, and two isochoric branches denoted by $B\rightarrow C$ and $D\rightarrow A$, along which the
optical cavity with the constant Hamiltonian is weakly coupled to
the hot reservoir or the cold reservoir, respectively. We now describe the four consecutive steps of the proposed Otto engine cycle as follows.

(i) Unitary compression $A\rightarrow B$: The system is isolated from the heat reservoir and undergoes a unitary compression in time $\tau_{ch}$. The system’s Hamiltonian is changed from $H^{ca}(\omega_{c})=\omega_{c}(\hat{a}^{\dag}\hat{a}+1/2)$ to $H^{ca}(\omega_{h})=\omega_{h}(\hat{a}^{\dag}\hat{a}+1/2)$ with a driving function $\omega_{ch}(t)=\omega_{c}\omega_{h}\tau_{ch}/[(\omega_{c}-\omega_{h})t+\omega_{h}\tau_{ch}]$, where $\hat{a}$ and $\hat{a}^{\dag}$ are the annihilation and creation operators of the oscillator, respectively. Because no heat is exchanged, the change in the system’s internal energy is equal to the work done on the system. We use a two-projective-measurement scheme \cite{Lutz20} to calculate the probability distribution of the quantum work as follows:
\begin{equation}\label{pwch}
p({w}_{ch})=\mathop{\sum}\limits_{n,m}\delta[{w_{ch}}-(\varepsilon_{m}^{h}-\varepsilon_n^{c})]p_{n\rightarrow m}^{{ch}}p_n^A,
\end{equation}
where $\varepsilon_n^{c}$ and $\varepsilon_{m}^{h}$ are the measured energies at the beginning and the end of this stroke, respectively. Here, $p_n^A$ is the initial occupation probability and $ p_{n\rightarrow m}^{{ch}}=|\langle n|U_{ch}|m\rangle|^2$ with the unitary operator $U_{ch}$ is the transition probability from 
state $|n\rangle$ to $|m\rangle$.

(ii) Isochoric heating $B\rightarrow C$: For this stroke, we consider the case in which only one atom of the correlated pair with the constant inverse temperature $\beta_h$ is sent to pass through the cavity. The Hamiltonian of the single radiation mode in the cavity is constant, with a fixed frequency $\omega_{h}$. Therefore, the stochastic heat injection is equivalent to an increase in the system eigenenergy, with no work being produced along the stroke.  {The probability distribution for
the heat absorbed by the system during this stroke, given that the stochastic compression work is $w_{ch}=\varepsilon_m^h-\varepsilon_n^c$, can then be expressed as \cite{Lutz20, jiao21}}
\begin{equation}\label{pqh}
  p({q}_{h}|{w}_{ch})=\mathop{\sum}\limits_{k,l}\delta[q_{h}-(\varepsilon_{l}^{h}-\varepsilon_{k}^{h})]p_{k\rightarrow l}^{_{h}}p_{k}^{B},
\end{equation}
where $\varepsilon_k^{h}$ and $\varepsilon_{l}^{h}$ are the measured energies at the beginning and the end of the heating stroke, respectively. $p_k^{B}=\delta_{km}$ indicates the probability that the system is in eigenstate $|m\rangle$ after the second projective measurement during the unitary compression, {and $p_{k\rightarrow l}^{{h}}$ is 
the probability that the system will collapse into the state $|l\rangle$ after the
second projective measurement for this hot isochore. In the case where the system reaches the unique steady state at the end of the isochore \cite{Lutz20,jiao21}, the probability satisfies the following generalized canonical form: $p^h_{k\rightarrow l}=e^{-\beta_h^{\mathrm{eff}} H^{ca}(\omega_h)
}/\mathrm{Tr}[e^{-\beta_h^{\mathrm{eff}} H^{ca}(\omega_h)}]$, where $\beta_h^{\mathrm{eff}}$ is called the effective temperature and will be defined in Eq. (\ref{beta}).}
   
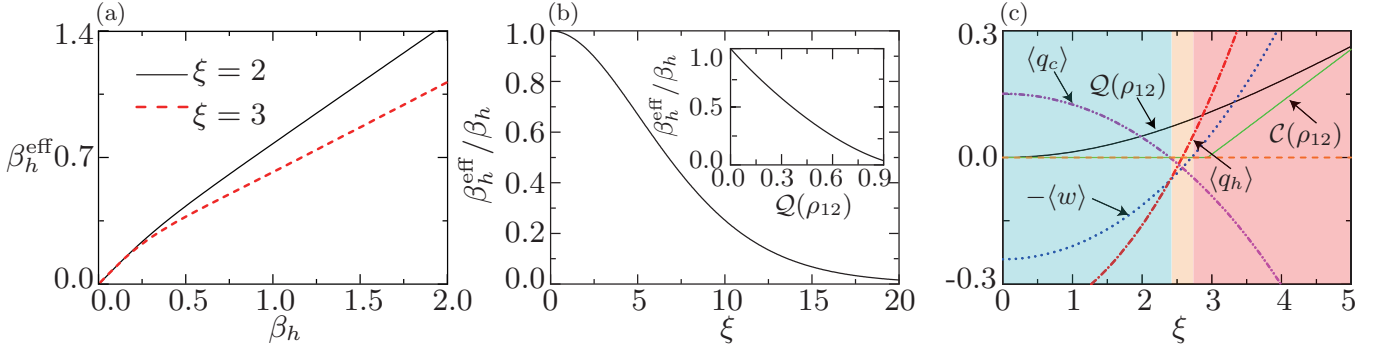
\begin{figure*} 
\begin{overpic}[width=5.9cm]{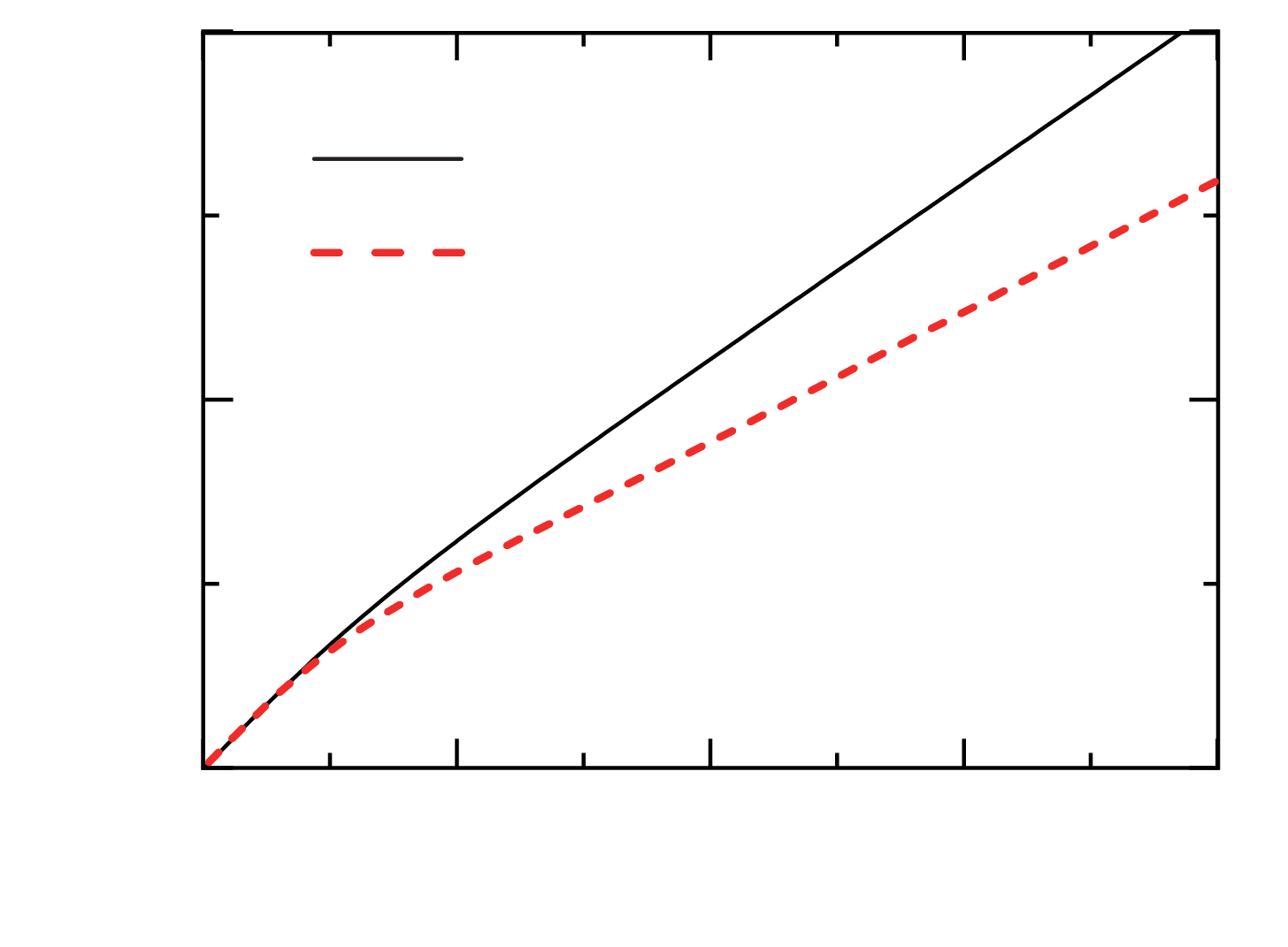}
    \put(15,73.5){(a)}
    \put(5.7,68){\scalebox{1.3}{1.4}}
    \put(5.7,40.5){\scalebox{1.3}{0.7}}
    \put(5.7,13){\scalebox{1.3}{0.0}}
    \put(14,8){\scalebox{1.3}{0.0}}
    \put(31.5,8){\scalebox{1.3}{0.5}}
    \put(51,8){\scalebox{1.3}{1.0}}
    \put(71,8){\scalebox{1.3}{1.5}}
    \put(91,8){\scalebox{1.3}{2.0}}
     \put(54,2){\scalebox{1.3}{$\beta_h$}}
     \put(37,60){\scalebox{1.3}{$\xi=2$}}
     \put(37,50){\scalebox{1.3}{$\xi=3$}}
     \put(-5,40.5){\scalebox{1.3}{$\beta_h^\mathrm{eff}$}}
\end{overpic}
\begin{overpic}[width=5.9cm]{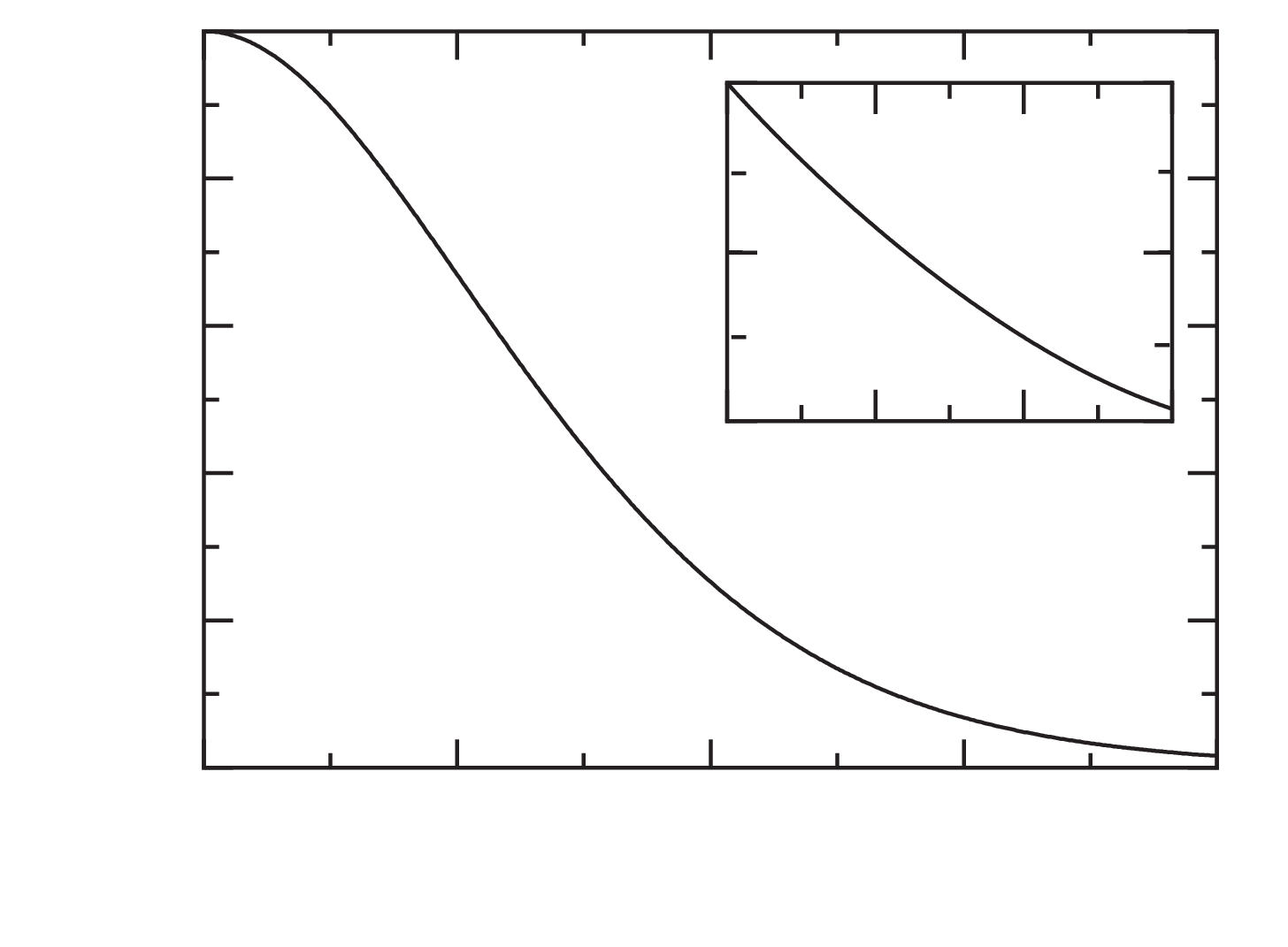}
    \put(15,73.5){(b)}
    \put(5.5,68){\scalebox{1.3}{1.0}}
    \put(5.5,57){\scalebox{1.3}{0.8}}
    \put(5.5,46){\scalebox{1.3}{0.6}}
    \put(5.5,34.5){\scalebox{1.3}{0.4}}
    \put(5.5,23.5){\scalebox{1.3}{0.2}}
    \put(5.5,13){\scalebox{1.3}{0.0}}
    \put(14,8){\scalebox{1.3}{0}}
    \put(34,8){\scalebox{1.3}{5}}
    \put(52,8){\scalebox{1.3}{10}}
    \put(71.5,8){\scalebox{1.3}{15}}
    \put(91,8){\scalebox{1.3}{20}}
    \put(54,2){\scalebox{1.3}{$\xi$}}
    \put(-5,32){\scalebox{1.3}{\rotatebox{90}{$\beta_h^\mathrm{eff}/\beta_h$}}}
    \put(38,47){\scalebox{1.1}{\rotatebox{90}{$\beta_h^\mathrm{eff}/\beta_h$}}}
    \put(47,63){\scalebox{1.1}{1.0}}
    \put(47,52){\scalebox{1.1}{0.5}}
    \put(47,40){\scalebox{1.1}{0.0}}
    \put(52,36){\scalebox{1.1}{0.0}}
    \put(63,36){\scalebox{1.1}{0.3}}
    \put(74,36){\scalebox{1.1}{0.6}}
    \put(85,36){\scalebox{1.1}{0.9}}
    \put(66,30){\scalebox{1.1}{$\mathcal{Q}(\rho_{12})$}}
\end{overpic}
\begin{overpic}[width=5.9cm]{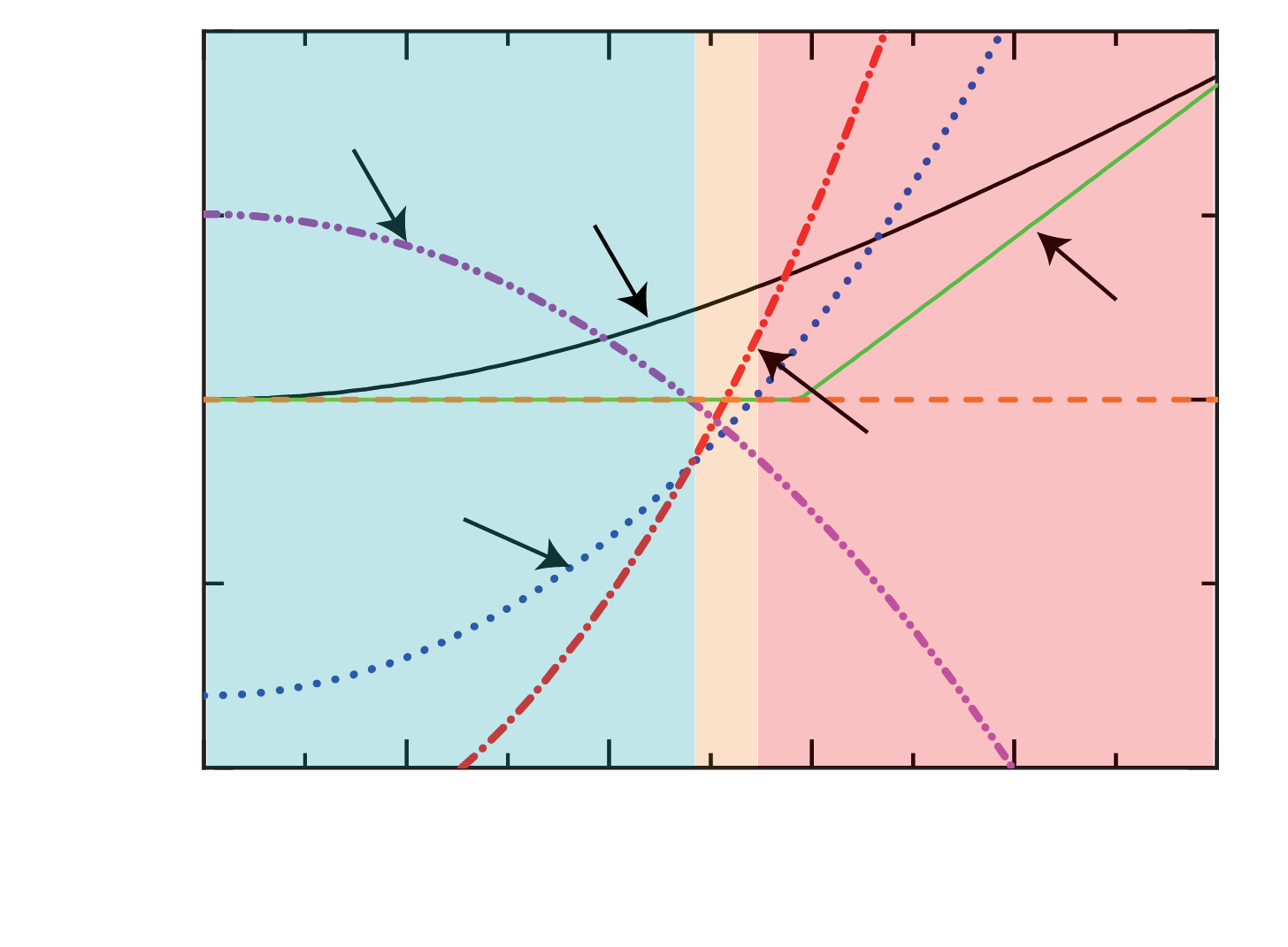}
    \put(15,73.5){(c)}
    \put(5.7,68){\scalebox{1.3}{0.3}}
    \put(5.7,40.5){\scalebox{1.3}{0.0}}
    \put(3,13){\scalebox{1.3}{-0.3}}
    \put(14,8){\scalebox{1.3}{0}}
    \put(30,8){\scalebox{1.3}{1}}
    \put(46,8){\scalebox{1.3}{2}}
    \put(61.5,8){\scalebox{1.3}{3}}
    \put(77.5,8){\scalebox{1.3}{4}}
    \put(92,8){\scalebox{1.3}{5}}
    \put(54,2){\scalebox{1.3}{$\xi$}}
    \put(21,63){\scalebox{1.1}{$\langle q_c \rangle$}}
    \put(62,36){\scalebox{1.1}{$\langle q_h \rangle$}}
    \put(21,32){\scalebox{1.1}{$-\langle w \rangle$}}
    \put(35,57){\scalebox{1.1}{$\mathcal{Q}(\rho_{12})$}}
    \put(76,46){\scalebox{1.1}{$\mathcal{C}(\rho_{12})$}}
            
\end{overpic}
{\caption{ (a) Effective temperature $\beta_h^\mathrm{eff}$ as a function of $\beta_h$ with $\omega_h=6$. (b) Dimensionless temperature $\beta_h^\mathrm{eff}/\beta_h$ as a function of the interaction strength $\xi$; the inset shows the dimensionless temperature $\beta_h^\mathrm{eff}/\beta_h$ as a function of the quantum discord $\mathcal{Q}(\rho_{12})$; the parameter $\beta_h=0.3$. (c) Work output $-\langle w\rangle$, quantum discord $\mathcal{Q}(\rho_{12})$, concurrence $C(\rho_{12})$, and the heats $\langle q_h\rangle $ and $\langle q_c\rangle$ absorbed by the system as functions of $\xi$. In (c) the parameters are $\beta_{c}=0.6$, $\beta_{h}=0.3$, $\omega_{h}=4.5$, $\omega_{c}=2$, and $\tau_\mathrm{dri}=0.95.$}\label{beff}}
\end{figure*}

The atom that passes through the optical cavity interacts with the cavity via a resonant Jaynes-Cummings coupling, which is given by $H^{int}=-\gamma(\hat{a}\sigma^{+}+\hat{a}^{\dag}\sigma^{-})$ \cite{Nar80} with a coupling constant $\gamma$. When an ensemble composed of many atoms is sent to the optical cavity, the cavity state dynamics can be described by \cite{Lutz09,Sch01}
\begin{eqnarray}\label{master}
\frac{d\rho_{t}}{dt}&=&{{i}[H^{ca}(\omega_h), \rho_t]}+r_{1}^{h}(\gamma\tau)^2(\hat{a}^\dag\rho_t\hat{a}-\frac{1}{2}\hat{a}\hat{a}^\dag\rho_t-\frac{1}{2}\rho_t\hat{a}\hat{a}^\dag)\nonumber\\
&+&r_{2}^{h}(\gamma\tau)^2(\hat{a}\rho_t\hat{a}^\dag-\frac{1}{2}\hat{a}^\dag \hat{a}\rho_t-\frac{1}{2}\rho_{t}\hat{a}^\dag \hat{a}),
\end{eqnarray}
where $\tau$ is the time spent by the atoms inside the cavity and $H^{ca}(\omega_h)=\omega_h (\hat{a}^\dag\hat{a}+\frac{1}2)$ is the Hamiltonian of the single-mode cavity. The coefficients $r_1^{h}$ and $r_2^{h}$ are the arrival rates for the atoms
in the excited and ground states, respectively, and these coefficients are thus associated with the probabilities of emission and absorption of a photon in the cavity, respectively. These two coefficients are determined using
$r_{1}^{h}=\rho_{e}^{h}+\rho_{d}^{h}$ and $r_2^{h}=\rho_{g}^{h}+\rho_{d}^h$, where $\rho_{e}^{h}=\mathrm{exp}(-\beta_{h}\omega_h)/[2\mathrm{cosh}(\beta_{h}\omega_{h})+2\mathrm{cosh}(\beta_{h}\xi)]$, $\rho_{g}^{h}=\mathrm{exp}(\beta_{h}\omega_{h})/[2\mathrm{cosh}(\beta_{h}\omega_{h})+2\mathrm{cosh}(\beta_{h}\xi)]$, and $\rho_{d}^{h}=\mathrm{cosh}(\beta_{h}\xi)/[2\mathrm{cosh}(\beta_{h}\omega_{h})+2\mathrm{cosh}(\beta_{h}\xi)]$. 

The optical cavity, as the working substance, is allowed to relax to the
stationary state at the end of the hot isochoric stroke, {where the first term on the right hand side of Eq. (\ref{master}) is vanishing because $[H^{ca}(\omega_h), {\rho_t}]=0$}, and the time duration $\tau_h$ of this stroke is assumed to be much shorter than the duration of the unitary stroke. 
The asymptotic steady-state solution to Eq. (\ref{master}), which describes the stationary state of the optical cavity, is then obtained as:
\begin{equation}\label{rhoss}
\rho^{ss}=(e^{\beta_{h}^\mathrm{eff} \omega_{h}/2}-e^{-\beta_{h}^\mathrm{eff} \omega_{h}/2})e^{-\beta_{h}^\mathrm{eff}H^{ca}(\omega_h)},
\end{equation}
where $\beta_{h}^\mathrm{eff}$ is introduced to denote the effective inverse temperature of the optical cavity. The detailed balance condition $\mathrm{exp}(-\beta_{h}^\mathrm{eff}\omega_h)=r_1^{h}/r_2^{h}$ then produces
 \begin{equation}
   {\beta_{h}^\mathrm{eff}}=\beta_h-\frac{1}{\omega_{h}}\mathrm{ln}\frac{1+e^{\beta_{h}\omega_{h}}\mathrm{cosh}(\beta_{h}\xi)}{e^{\beta_{h}\omega_{h}}+\mathrm{cosh}(\beta_{h}\xi)}. \label{beta}
 \end{equation}

 {Given the frequency $\omega_h$ and the interaction strength $\xi$, the effective inverse temperature $\beta_h^{\mathrm{eff}}$ will thus increase monotonically with increasing inverse temperature $\beta_h$, as illustrated in Fig. \ref{beff}(a).
The normalized temperature $\beta_h^{\mathrm{eff}}/\beta_h$ as a function of $\xi$ is plotted in Fig. \ref{beff}(b), demonstrating that $\beta_h^{\mathrm{eff}}/\beta_h$ is a monotonically decreasing function of $\xi$ and $\beta_h^{\mathrm{eff}}/\beta=1$ for a vanishing $\xi$, as expected. The discord $\mathcal{Q}(\rho_{12})$ given by Eq. (\ref{Qd}) is determined by the interaction strength $\xi$, which means that the effective temperature $\beta_h^{\mathrm{eff}}$ is closely dependent on the
 discord $\mathcal{Q}(\rho_{12})$. Because the discord is a monotonically increasing function of the interaction strength, the ratio $\beta_h^{\mathrm{eff}}/\beta_h$ decreases with increasing $\mathcal{Q}(\rho_{12})$ [see the inset of Fig. \ref{beff}(b)]. Physically, a greater discord means that the system deviates further away from the thermal state of the inverse temperature $\beta$. We also note from Eq. (\ref{beta}) that
 the expression for the normalized temperature $\beta_h^{\mathrm{eff}}/\beta_h$ 
 can be simplified to give
$ {\beta_{h}^\mathrm{eff}}/{\beta_{h}}\simeq1-(\beta_{h}\xi)^2/4=1-2\mathcal{Q}(\rho_{12})\mathrm{ln}2$
in the high-temperature and/or weak-coupling limit }.

(iii) Unitary expansion $C \rightarrow D$: The optical cavity is isolated again from the heat reservoir and undergoes a unitary expansion during the time period $\tau_{hc}$ with the driving function $\omega_{hc}(t)=\omega_{c}\omega_{h}\tau_{hc}/[(\omega_{h}-\omega_{c})(t-\tau_{ch})+\omega_c\tau_{hc}]$.
Because this stroke can be accomplished by reversing the protocol used in the
compression process described above, we set $\tau_{\mathrm{dri}}\equiv \tau_{hc}=\tau_{ch} $ to obtain $\omega_{hc}(t)=\omega_{ch}(2\tau_{\mathrm{dri}}-t)$. During this stroke, no heat is exchanged, and the change in the system’s internal energy is
equal to the work done on the system. {For the given quantum expansion work $w_{ch}$ and heat injection $q_h$,  by using the two-point-measurement scheme again, the probability distribution of the stochastic work during the expansion process is given by} 
\begin{equation}\label{pwhc}
p({w_{hc}}|{w_{ch}},{q}_{h})=\mathop{\sum}\limits_{i,j}\delta[{w_{hc}}-(\varepsilon_j^c-\varepsilon_i^h)]p_{i\rightarrow j}^{{hc}}p_i^{C},
\end{equation}
where $p_i^{C}=\delta_{il}$ is the probability that the optical cavity is in the eigenstate $|l\rangle$ after the second  projective measurement in the isochoric heating. In addition, $p_{i\rightarrow j}^{hc}=|\langle j|U_{hc}|i\rangle|^2$ is the transition probability between the instantaneous eigenstates $|i\rangle$ and $|j\rangle$, where $U_{hc}$ is the unitary operator.

(iv) Isochoric cooling $D\rightarrow A$: The system, which has the constant Hamiltonian  $H^{ca}(\omega)=H^{ca}(\omega_c)$, is coupled to a thermal reservoir of constant inverse temperature $\beta_{c}$ in the time duration $\tau_c$, which is much shorter than the adiabatic driving time $\tau_{\mathrm{dri}}$. The system reaches thermal equilibrium with the heat reservoir at the end point of the cold isochore, and the state of the system at this end point is given by $\rho_{0}=e^{-\beta_{c}H^{ca}(\omega_{c})}/\mathrm{Tr}[e^{-\beta_{c}H^{ca}(\omega_{c})}]$.

After a single cycle, the joint distribution of the total work ${w}$ and the heat injection ${q}_{h}$ can be calculated by combining Eqs. (\ref{pwch}), (\ref{pqh}), and (\ref{pwhc}) to give \cite{Lutz20,Hol121,Lutz21,jiao21}
\begin{eqnarray}\label{JP}
p({w},{q_{h}})&=&\mathop{\sum}\limits_{n,m,i,j}\delta({w}+\varepsilon_n^{c}-\varepsilon_j^{c}+\varepsilon_i^{h}-\varepsilon_m^{h})\nonumber\\
&\times&\delta({q_h}-\varepsilon_{i}^{h}+\varepsilon_{m}^{h})
|\langle m|U_{ch}|n\rangle|^2|\langle j|U_{hc}|i\rangle|^2\nonumber\\
&\times&\frac{e^{-\beta_{c}\varepsilon_{n}^{c}}e^{-\beta_{h}^\mathrm{eff}\varepsilon_{i}^{h}}}{Z_{c}^{ca}Z_{h}^{ca}},
\end{eqnarray}
where the partition functions of the optical cavity are $Z_{c}^{ca}=\mathrm{Tr}[e^{-\beta_{c}H^{ca}(\omega_{c})}]$ and $Z_{h}^{ca}=\mathrm{Tr}[e^{-\beta_{h}^\mathrm{eff}H^{ca}(\omega_{h})}]$.

\subsection{Machine performance and fluctuations}\label{QOm}

To determine the statistics of the work and the heat, the characteristic function of the joint distribution function given in Eq. (\ref{JP}), denoted by $G(u,v)\equiv\int\int p(w,q_h)e^{-i v q_h-i u w}dwdq_h$, can be obtained as follows \cite{Lutz21,Fei22}:
\begin{equation}\label{cf}
G(u,v)=\langle  e^{-iv{q_h}-iuw}\rangle=\frac{G_c G_h}{1+\phi},
\end{equation}
where
\begin{equation}\label{Gc}
   G_c=\frac{e^{\beta_{c}\omega_{c}}-1}{\sqrt{r_\phi \mathrm{cosh}(\beta_{c}\omega_{c}+
i v_0)+\mathrm{cosh}(\beta_{c}\omega_{c}+ i u_0)-2}}, \end{equation}
and
\begin{equation}\label{Gh}
G_h=\frac{e^{\beta_{h}^\mathrm{eff}\omega_{h}}-1}{\sqrt{r_\phi\mathrm{cos}(v_0-i\beta_{h}^\mathrm{eff}\omega_{h})+\mathrm{cos}(i\beta_{h}^\mathrm{eff}\omega_{h}+u_0)-2}}.
\end{equation}
Here, we have used $u_0=u(\omega_c-\omega_h)+v\omega_h$, $v_0=u(\omega_c+\omega_h)-v\omega_h$, and $r_\phi=(1-\phi)/(1+\phi)$. {As shown in Appendix \ref{aped}, the parameter $\phi$ can be derived to be 
\begin{equation} \label{phi1}
\phi\equiv 1+\frac{1-\cosh\left(\sqrt{1-\zeta}\ln\frac{\omega_h}{\omega_c}\right)}{\zeta-1},
\end{equation}
where $\zeta=[2\tau_{\mathrm{dri}}\omega_{c}\omega_{h}/(\omega_{h}-\omega_{c})]^2$}, and this parameter is called the nonadiabatic factor. {In the case where $\zeta>1$, Eq. (\ref{phi1}) then becomes
$  
 \phi=  1+\frac{1-\mathrm{cos}[\sqrt{\zeta-1}\mathrm{ln}(\omega_{h}/\omega_{c})]}{\zeta-1} $, indicating that the extreme limit of $\zeta
 \gg 1$ leads to $\phi\rightarrow 1$. Within the long time limit, $\phi$ approaches $1$, and thus the quantum adiabatic condition is satisfied }. 
 \begin{figure*}
    \centering
    \begin{overpic}[width=5.9cm]{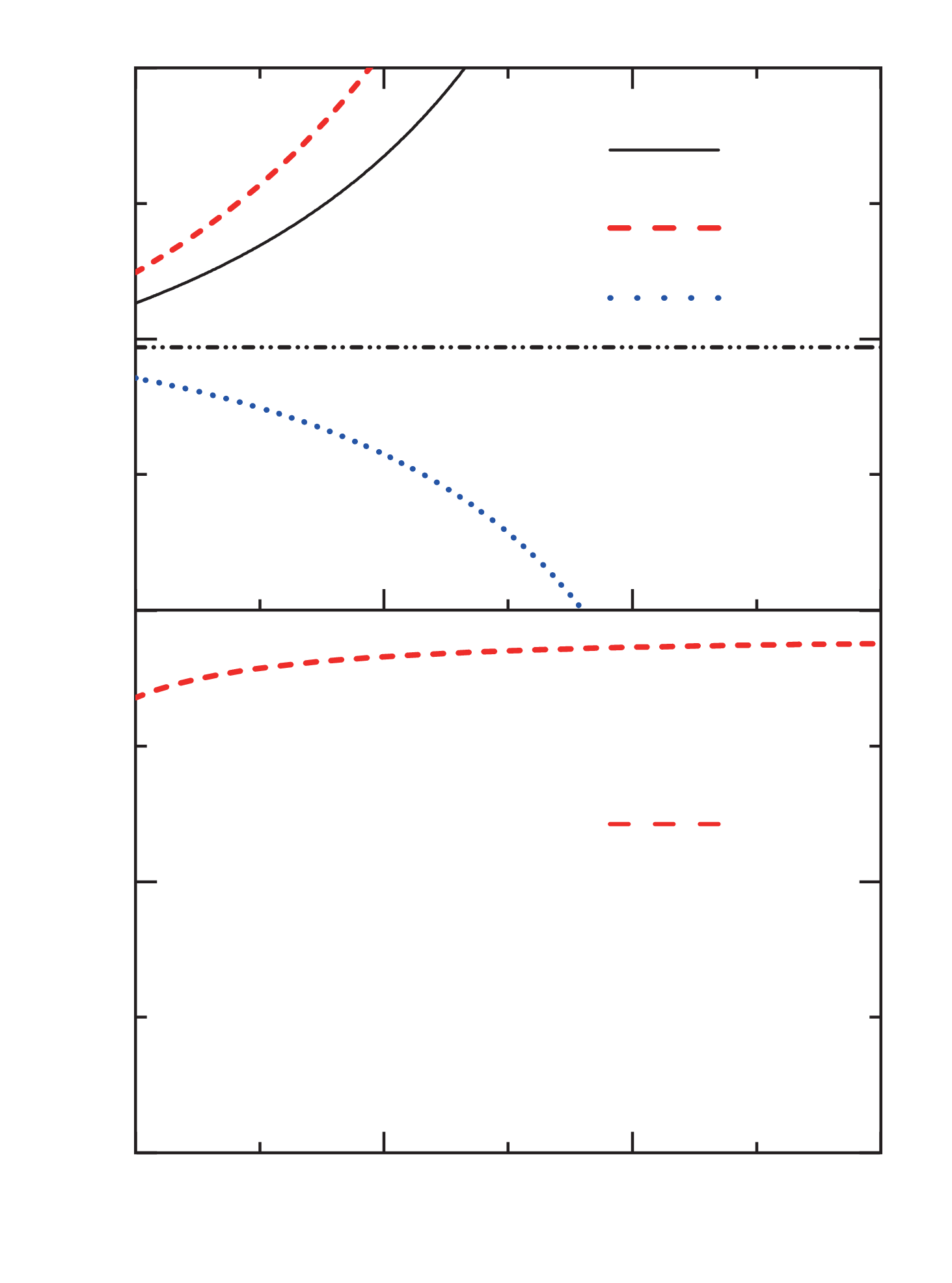}
    \put(9.5,96.5){(a)}
     \put(15,96.5){\scalebox{1.3}{$\beta_h=0.15$}}
    \put(6.5,92){\scalebox{1.3}{5}}
    \put(6.5,72.5){\scalebox{1.3}{0}}
    \put(2.5,53.5){\scalebox{1.3}{$-5$}}
    \put(-0.2,49.5){\scalebox{1.3}{0.70}}
    \put(-0.2,30){\scalebox{1.3}{0.35}}
    \put(-0.2,9.5){\scalebox{1.3}{$0.00$}}
    \put(6.5,5.5){\scalebox{1.3}{0.0}}
\put(26,5.5){\scalebox{1.3}{0.3}}
\put(46,5.5){\scalebox{1.3}{0.6}}
\put(66,5.5){\scalebox{1.3}{0.9}}
\put(33,1){\scalebox{1.3}{$\mathcal{Q}(\rho_{12})$}}
\put(56,87){\scalebox{1.1}{$-\langle w\rangle$}}
\put(57,81){\scalebox{1.1}{$\langle q_h \rangle$}}
\put(57,75.5){\scalebox{1.1}{$\langle q_c \rangle$}}
\put(57,34.5){\scalebox{1.3}{$\eta_{th}$ }}
\end{overpic}
\begin{overpic}[width=5.9cm]{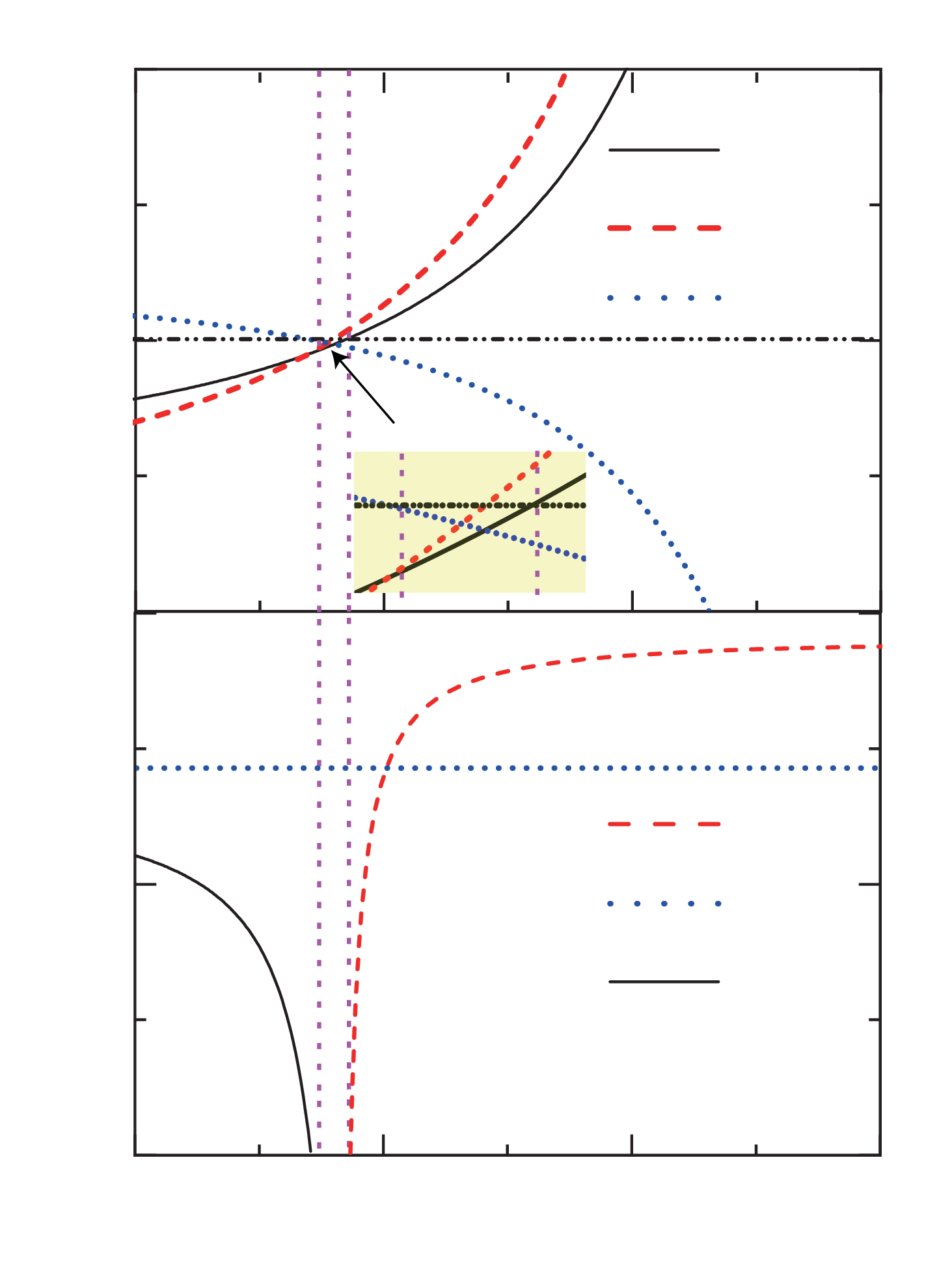}
    \put(9.5,96.5){(b)}
    \put(15,96.5){\scalebox{1.3}{$\beta_h=0.3$}}
    \put(6.5,92){\scalebox{1.3}{5}}
    \put(6.5,72.5){\scalebox{1.3}{0}}
    \put(2.5,53.5){\scalebox{1.3}{$-5$}}
   \put(-0.2,49.5){\scalebox{1.3}{0.70}}
    \put(-0.2,30){\scalebox{1.3}{0.35}}
    \put(-0.2,9.5){\scalebox{1.3}{$0.00$}}
    \put(6.5,5.5){\scalebox{1.3}{0.0}}
    \put(26,5.5){\scalebox{1.3}{0.3}}
    \put(46,5.5){\scalebox{1.3}{0.6}}
    \put(66,5.5){\scalebox{1.3}{0.9}}
    \put(33,1){\scalebox{1.3}{$\mathcal{Q}(\rho_{12})$}}
    \put(56,87){\scalebox{1.2}{$-\langle w\rangle$}}
\put(57,81.5){\scalebox{1.1}{$\langle q_h \rangle$}}
\put(57,75.5){\scalebox{1.1}{$\langle q_c \rangle$}}
\put(57,34.5){\scalebox{1.3}{$\eta_{th}$ }}
\put(57,28.5){\scalebox{1.3}{$\eta_{C}$ }}
\put(57,22.5){\scalebox{1.3}{$\varepsilon$ }}
\put(25,80){\rotatebox{90}{\scalebox{0.9}{Heater}}}
\put(25,80){\rotatebox{90}{\scalebox{0.9}{Heater}}}
\put(16,75.5){\rotatebox{90}{\scalebox{0.9}{refrigerator}}}
\put(16,75.5){\rotatebox{90}{\scalebox{0.9}{refrigerator}}}
\put(46,67){\scalebox{0.9}{Heat engine}}
\put(46,67){\scalebox{0.9}{Heat engine}}
\end{overpic}
\begin{overpic}[width=5.9cm]{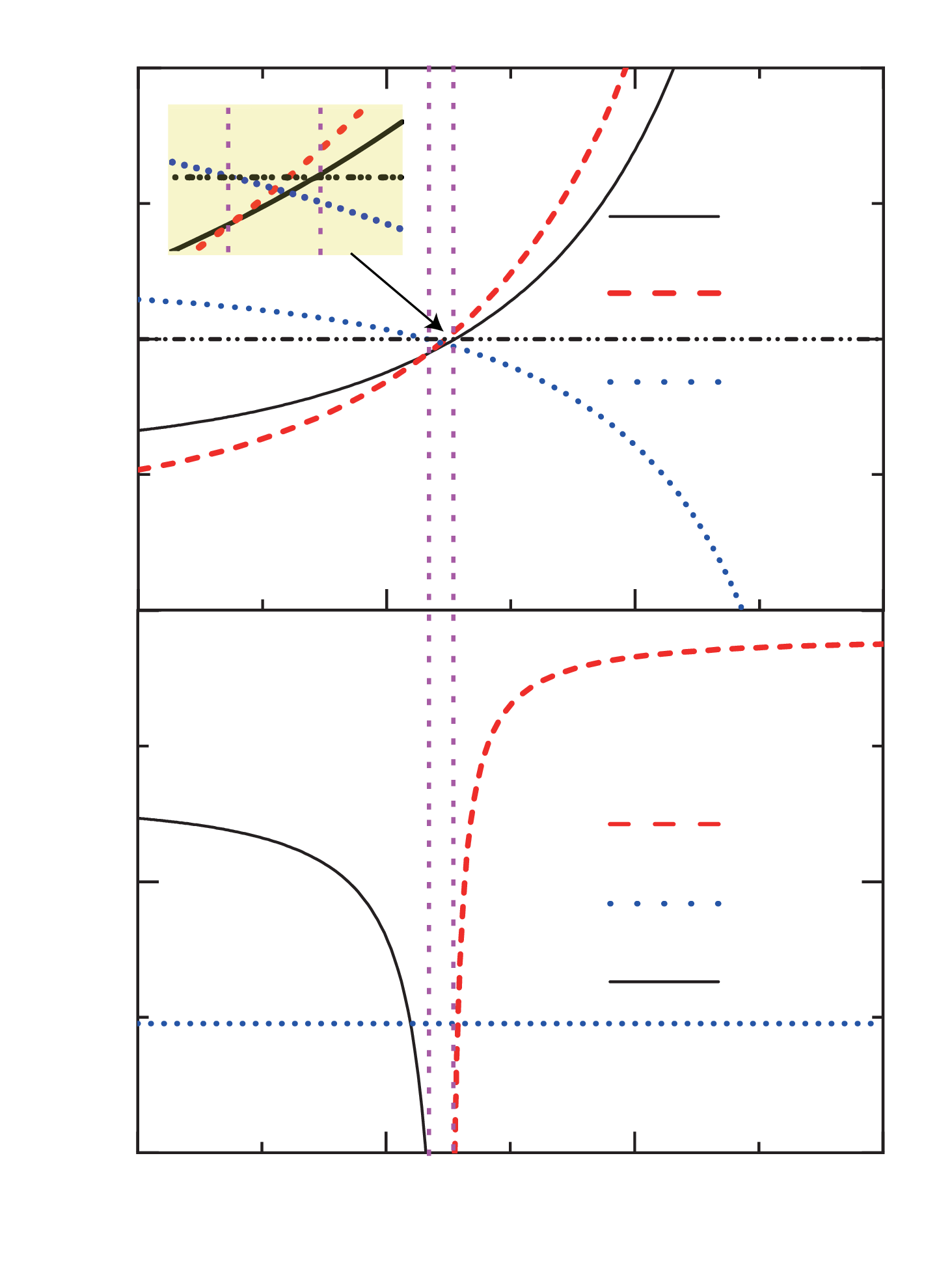}
    \put(9.5,96.5){(c)}
    \put(15,96.5){\scalebox{1.3}{$\beta_h=0.5$}}
    \put(6.5,92){\scalebox{1.3}{5}}
    \put(6.5,72.5){\scalebox{1.3}{0}}
    \put(2.5,53.5){\scalebox{1.3}{$-5$}}
    \put(-0.2,49.5){\scalebox{1.3}{0.70}}
    \put(-0.2,30){\scalebox{1.3}{0.35}}
    \put(-0.2,9.5){\scalebox{1.3}{$0.00$}}
    \put(6.5,5.5){\scalebox{1.3}{0.0}}
    \put(26,5.5){\scalebox{1.3}{0.3}}
    \put(46,5.5){\scalebox{1.3}{0.6}}
    \put(66,5.5){\scalebox{1.3}{0.9}}
    \put(33,1){\scalebox{1.3}{$\mathcal{Q}(\rho_{12})$}}
    \put(56,82){\scalebox{1.1}{$-\langle w\rangle$}}
\put(57,76){\scalebox{1.1}{$\langle q_h \rangle$}}
\put(57,69){\scalebox{1.1}{$\langle q_c \rangle$}}
\put(57,34.5){\scalebox{1.3}{$\eta_{th}$ }}
\put(57,28.5){\scalebox{1.3}{$\eta_{C}$ }}
\put(57,22.5){\scalebox{1.3}{$\varepsilon$ }}
\end{overpic}
    \caption{{Work output $-\langle w\rangle $ and heat values $\langle q_{h,c}\rangle $ absorbed by the system (upper panel), and the thermodynamic efficiency $\eta_{th}$ (lower panel) as a function of the quantum discord $\mathcal{Q}(\rho_{12})$ for (a) $\beta_h=0.15$, (b) $\beta_h=0.3$, and (c)$\beta_h=0.5$. The parameter values here are $\beta_{c}=0.6$, $\omega_{h}=6$, $\omega_{c}=2$, and the driving time $\tau_\mathrm{dri}=0.8$.}}
    \label{wqeta}
\end{figure*}
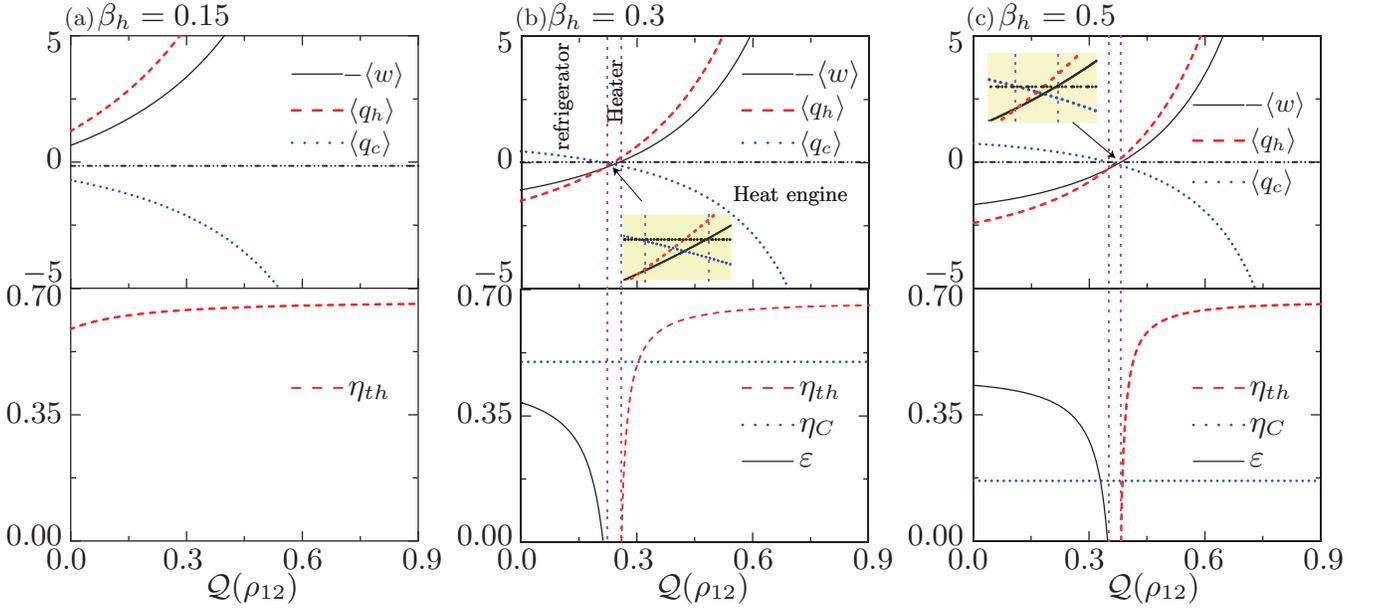

  The average and the variance of the work, and the average heat injection, can be
  obtained explicitly as
 \begin{eqnarray}\label{aw}
\langle{w}\rangle&=&-i\frac{\partial\mathrm{ln}G(u,v)}{\partial u}\bigg|_{u=v=0}\nonumber\\
&=&\omega_{h}(\phi\langle n_{c}^{eq}\rangle-\langle n_{h}^{ss}\rangle)+\omega_{c}(\phi\langle n_{h}^{ss}\rangle-\langle n_{c}^{eq}\rangle), 
\end{eqnarray}
\begin{eqnarray}\label{var}
\delta w^{2}&=&\langle w^2\rangle-\langle w\rangle^2
=-\frac{\partial^2\mathrm{ln}G(u,v)}{\partial u^2}\bigg|_{u=v=0}\nonumber\\
&=&\omega_{h}^2[-\frac{1}{2}+(2\phi^2-1)\langle n_{c}^{eq}\rangle^2+\langle n_{h}^{ss}\rangle^2]\nonumber\\
&+&\omega_{c}^2[-\frac{1}{2}+\langle n_{c}^{eq}\rangle^2+(2\phi^2-1)\langle n_{h}^{ss}\rangle^2]\\\nonumber
&+&\omega_{h}\omega_{c}\phi(1-2\langle n_{c}^{eq}\rangle^2-2\langle n_{h}^{ss}\rangle^2),
\end{eqnarray}
\begin{eqnarray}\label{aq}
\langle {q_h}\rangle=-i\frac{\partial\mathrm{ln}G(u,v)}{\partial v}\bigg|_{u=v=0}
=\omega_{h}(\langle n_{h}^{ss}\rangle-\phi\langle n_{c}^{eq}\rangle),
\end{eqnarray}
respectively, where we have used $\langle n_{c}^{eq}\rangle=\mathrm{coth}(\beta_{c}\omega_{c}/2)/2$ and $\langle n_{h}^{ss}\rangle= {\mathrm{coth}(\beta_{h}^\mathrm{eff}\omega_{h}/2)}/{2}$ to denote
 the mean numbers of the cavity with the cold thermal reservoir and the hot nonthermal reservoir, respectively.

The average values of the work and the heat injection depend on the quantum discord, which enters into $\langle n_h^{ss}\rangle$, and under this
condition, even the
operation mode can change when the quantum discord is involved. Notably,
the mean work given by Eq. (\ref{aw}) can be split into the following sum of two parts: $-\langle{w}\rangle=\langle {w_\mathrm{adi}}\rangle-\langle{w_\mathrm{fric}}\rangle$, where the first part $\langle w_\mathrm{adi}\rangle=(\omega_{h}-\omega_{c})(\langle n_{h}^{ss}\rangle-\langle 
 n_\mathrm{c}^{eq}\rangle)$ is the mean work in the quantum adiabatic case, and the second part $\langle{w_\mathrm{fric}}\rangle=(\phi-1)(\omega_{h}\langle n_{c}^{eq}\rangle+\omega_{c}\langle n_{h}^{ss}\rangle)$ is the frictional work caused by diabatic transitions occurring along the two unitary driven processes.

{As repeatedly emphasized in the work above, 
the quantum entanglement is vanishing in the case of multipartite separable states, where the quantum discord is, however, nonzero [see Fig. \ref{QC}(b)]. The quantum discord is thus considered to be a more general quantum resource
 than the quantum concurrence in our engine model}. {To complete
this picture, we have produced, as shown in Fig. \ref{beff}(c), the average values of the work, the absorbed heat, the concurrence, and the discord as functions of the interaction strength $\xi$. We see that, for the given parameters, $\mathcal{Q}(\rho_{12})>0$ for $\xi>0$ but $C(\rho_{12})>0$ only for $\xi>2.922$; however, in the range where $C(\rho_{12})$ is finite, both $C(\rho_{12})$ and $\mathcal{Q}(\rho_{12})$ are monotonically increasing functions of $\xi$. Furthermore, the ranges of the quantities ($-\langle w\rangle$ and $\langle q_{c,h}\rangle)$ when parametrized using $\mathcal{Q}(\rho_{12})$
are much larger than those obtained when using $C(\rho_{12})$. Therefore, we use the quantum discord rather than the concurrence in the analysis of the quantum machine’s performance.}

The thermodynamic efficiency, which acts as one of performance measures and is defined by $\eta_{th}=-\langle {w}\rangle/\langle q_h\rangle$, where $\langle {w}\rangle$ and $\langle {q_h}\rangle$ were given by Eqs. (\ref{aw}) and (\ref{aq}), respectively, can then be obtained as
\begin{equation}\label{eff}
\eta_{th}=1-\frac{\omega_c}{\omega_h}-\frac{\omega_{c}}{\omega_{h}}\frac{(\phi-1)(\langle n_{h}^{ss}\rangle+\langle n_{c}^{eq}\rangle)}{\langle n_{h}^{ss}\rangle-\phi\langle n_{c}^{eq}\rangle},
\end{equation}
which reduces to the so-called Otto efficiency, $\eta_O=1-\omega_\mathrm{c}/\omega_\mathrm{h}$ if two unitary strokes are quantum adiabatic, irrespective of the existence of any correlations between the two atoms.
  The efficiency (\ref{eff}) can then be rewritten as $\eta_{th}=\eta_{C}^\mathrm{gen}-\langle \sigma\rangle/(\beta_{h}^\mathrm{eff} \langle q_{h}\rangle)$, where $\eta_{C}^{\mathrm{gen}}=1-\beta_h^\mathrm{eff}/\beta_c$ is the so-called generalized Carnot efficiency, and $\langle\sigma\rangle=-\beta_{h}^\mathrm{eff}\langle{q}_{h}\rangle-\beta_{c}\langle{q}_{c}\rangle\ge 0$ is the average 
entropy production of the heat engine during a single cycle \cite{ar22}. Although the efficiency $\eta_{\mathrm{th}}$ may surpass the Carnot efficiency $\eta_{C}=1-\beta_{h}/\beta_{c}$, it does satisfy the second law of thermodynamics because it must be bound by the generalized Carnot value $\eta_C^{\mathrm{gen}}$ as a result of positive entropy production.   

{Figure \ref{wqeta}(a) shows that, for $\beta_c=0.6$ and $\beta_h=0.15$, as the quantum discord increases, both the mean heat injection $\langle {q_h}\rangle$ and the average work $-\langle w\rangle$ also increase, leading to higher thermodynamic efficiency $\eta_{th}$. Given the specifications selected for the control parameters, including $\omega_{c,h}$ and $\beta_{c,h}$, the machine always operates as a heat engine with $-\langle w\rangle>0$ and $\eta_{th}>0$ for finite or vanishing quantum discord. 
 However, when different bath temperature values $\beta_{c,h}$ are selected, the machine mode may be changed by varying the discord}. When the control parameters $\omega_{c,h}$ are given for appropriate values of the bath temperature, gradually increasing $\mathcal{Q}(\rho_{12})$ results in the occurrence of the following three consecutive operating modes [upper panels, Figs. \ref{wqeta}(b) and \ref{wqeta}(c) ]: (i) the refrigerator mode ($-\langle {w}\rangle<0, \langle {q_h}\rangle<0$, $\langle {q_c}\rangle>0$), (ii) the heater mode ($-\langle {w}\rangle<0, \langle{q_c}\rangle<0$), and (iii) the heat engine mode ($-\langle{w}\rangle>0$, $\langle {q_h}\rangle>0$, $\langle {q_c}\rangle<0$).

{When the quantum discord $\mathcal{Q}(\rho_{12})$ increases gradually, the thermodynamic efficiency also increases and tends towards the Otto limit $\eta_O=1-\omega_c/\omega_h$, as shown in lower panels of Fig. {\ref{wqeta}}(a)$-${\ref{wqeta}}(c).}
For the heater mode, where the heat must flow into the cold reservoir, $\langle
q_c\rangle< 0$, the average heat injection $\langle q_h\rangle$ can be an arbitrary real number, and no useful work
is extracted, i.e., $-\langle w\rangle \le 0$. When operating as a refrigerator, the machine absorbs
the heat from the cold reservoir, which entails $\langle q_c\rangle>0$, and releases this heat into the hot reservoir, such that $\langle q_h\rangle<0$, and work should be consumed, i.e., $-\langle w\rangle<0$. The refrigerator’s coefficient of performance is defined as the ratio of the cooling load to the work input, i.e., $\varepsilon=\langle q_c\rangle/\langle w\rangle$, which is positive [see the lower panels in Figs. \ref{wqeta}(b) and \ref{wqeta}(c)]. {We also note from comparison of Fig. \ref{wqeta}(b) with Fig. \ref{wqeta}(c) that, even at a constant quantum discord, the machine mode can be changed when the bath temperature is modified.}

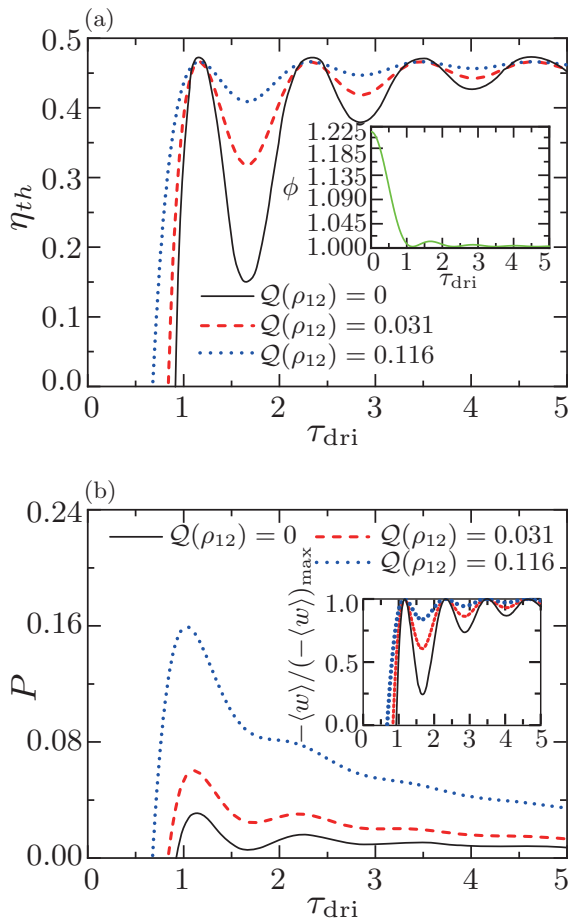
\begin{figure}
\begin{overpic}[width=8cm]{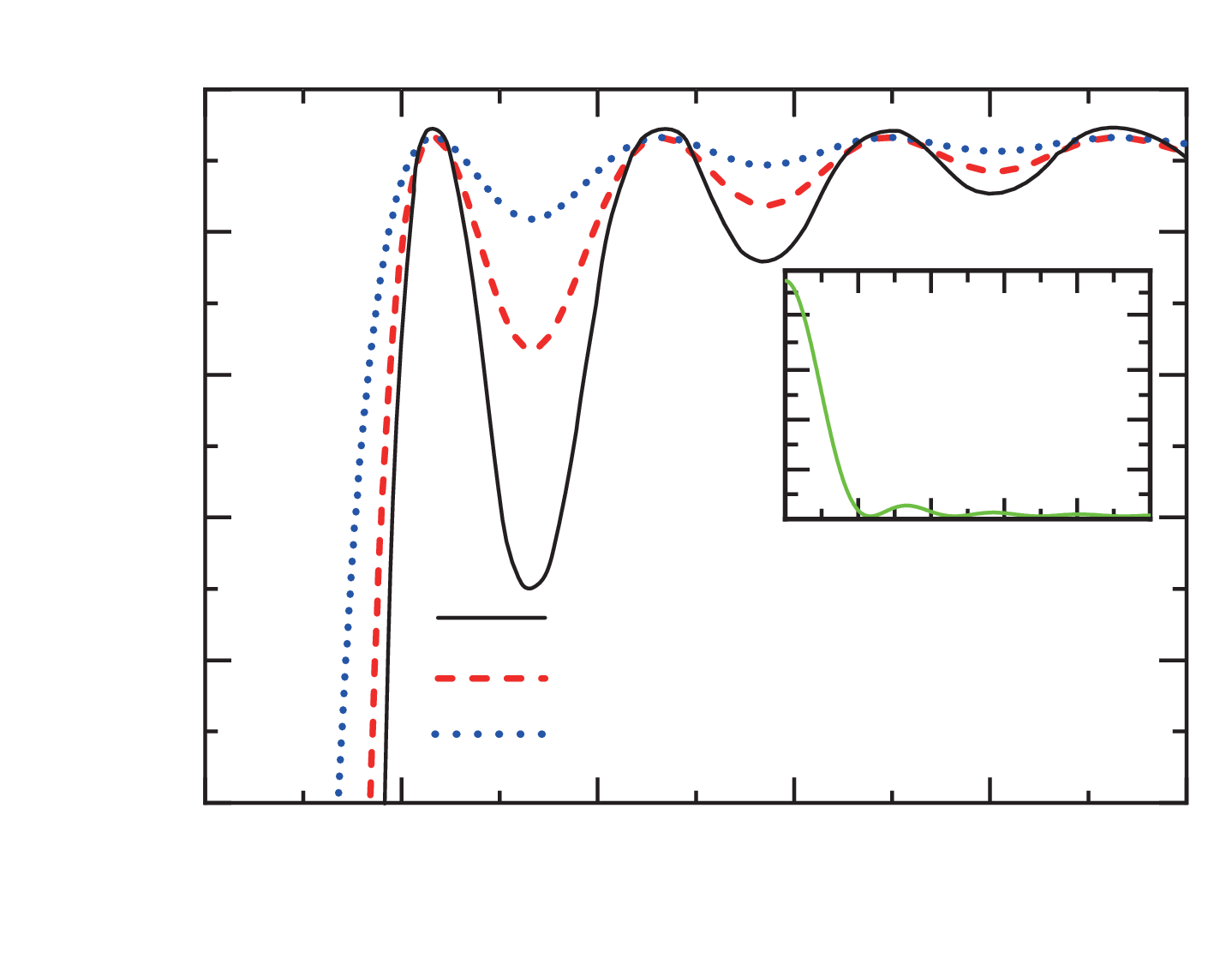}
    \put(16,73){(a)}
    \put(9,69){\scalebox{1.3}{0.5}}
    \put(9,57.5){\scalebox{1.3}{0.4}}
    \put(9,46){\scalebox{1.3}{0.3}}
    \put(9,34.5){\scalebox{1.3}{0.2}}
    \put(9,23){\scalebox{1.3}{0.1}}
    \put(9,12.5){\scalebox{1.3}{0.0}}
     \put(16,8){\scalebox{1.3}{0}}
    \put(31,8){\scalebox{1.3}{1}}
    \put(47,8){\scalebox{1.3}{2}}
    \put(63,8){\scalebox{1.3}{3}}
    \put(79,8){\scalebox{1.3}{4}}
    \put(94,8){\scalebox{1.3}{5}}
    \put(53,4){\scalebox{1.5}
    {$\tau_\mathrm{dri}$}}
    \put(63,32.5){\scalebox{1.1}{0}}
    \put(68.5,32.5){\scalebox{1.1}{1}}
    \put(74.5,32.5){\scalebox{1.1}{2}}
    \put(80.5,32.5){\scalebox{1.1}{3}}
    \put(86,32.5){\scalebox{1.1}{4}}
    \put(91.5,32.5){\scalebox{1.1}{5}}
    \put(53,54){\scalebox{1.1}{1.225}}
    \put(53,50.5){\scalebox{1.1}{1.185}}
    \put(53,46.5){\scalebox{1.1}{1.135}}
    \put(53,43){\scalebox{1.1}{1.090}}
    \put(53,39){\scalebox{1.1}{1.045}}
    \put(53,35){\scalebox{1.1}{1.000}}
    \put(49,45){\scalebox{1.1}{$\phi$}}
    \put(45,27){\scalebox{1.1}{$\mathcal{Q}(\rho_{12})=0$}}
    \put(45,22){\scalebox{1.1}{$\mathcal{Q}(\rho_{12})=0.031$}}
    \put(45,17){\scalebox{1.1}{$\mathcal{Q}(\rho_{12})=0.116$}}
    \put(75,30){\scalebox{1.1}{$\tau_\mathrm{dri}$}}
    \put(4,39){\rotatebox{90}{\scalebox{1.5}{$\eta_{th}$}}}
\end{overpic}
\begin{overpic}[width=8cm]{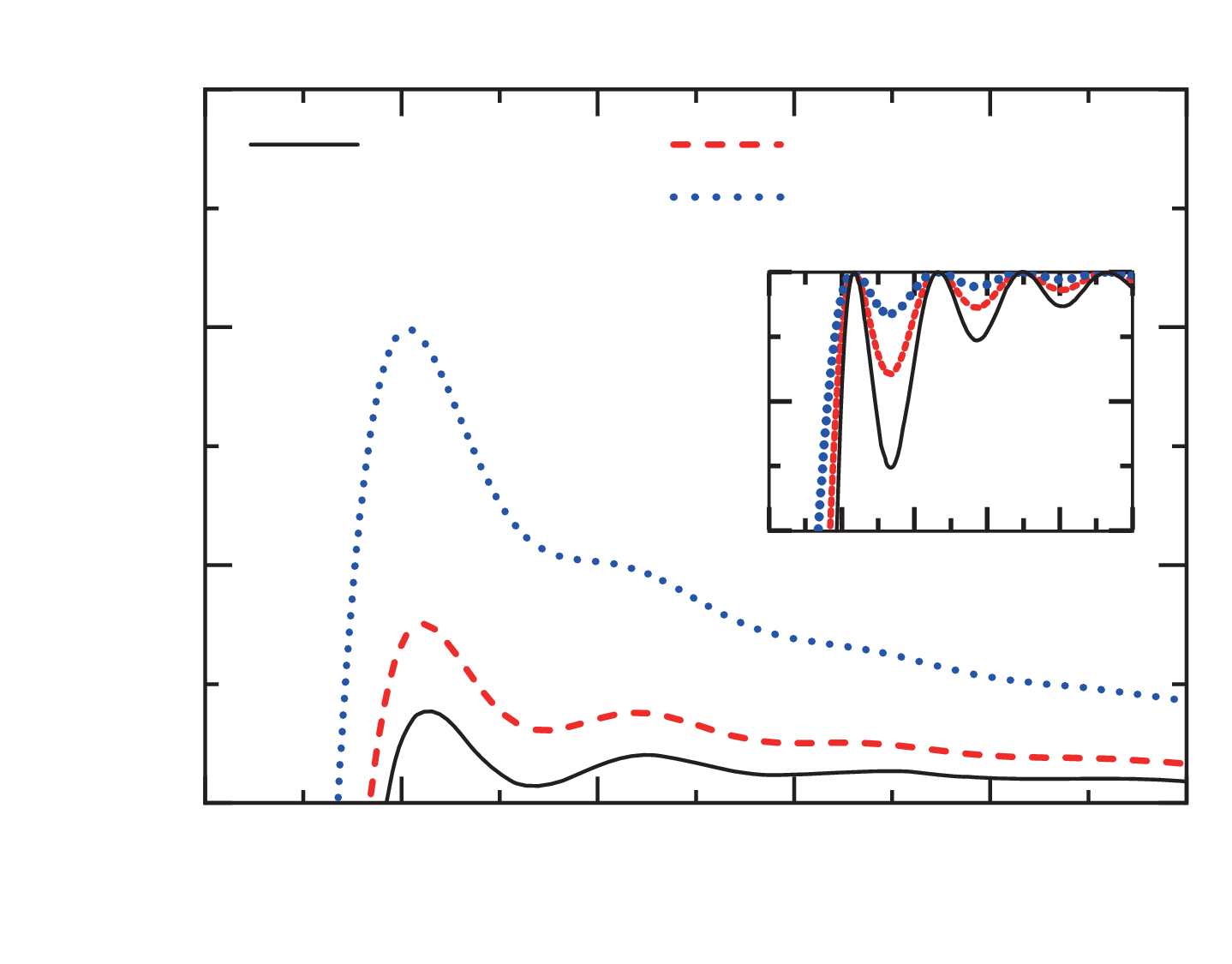}
    \put(16,73){(b)}
     \put(6.5,69){\scalebox{1.3}{0.24}}
    \put(6.5,50.5){\scalebox{1.3}{0.16}}
    \put(6.5,31){\scalebox{1.3}{0.08}}
    \put(6.5,12){\scalebox{1.3}{0.00}}
    \put(16,8){\scalebox{1.3}{0}}
    \put(31,8){\scalebox{1.3}{1}}
    \put(47,8){\scalebox{1.3}{2}}
    \put(63,8){\scalebox{1.3}{3}}
    \put(79,8){\scalebox{1.3}{4}}
    \put(94,8){\scalebox{1.3}{5}}
    \put(53,4){\scalebox{1.5}{$\tau_\mathrm{dri}$}}
    \put(4,39){\rotatebox{90}{\scalebox{1.5}{$P$}}}
    \put(30,65.5){\scalebox{1.1}{$\mathcal{Q}(\rho_{12})=0$}}
    \put(65,65.5){\scalebox{1.1}{$\mathcal{Q}(\rho_{12})=0.031$}}
    \put(65,61){\scalebox{1.1}{$\mathcal{Q}(\rho_{12})=0.116$}}
    \put(56,54){\scalebox{1.1}{1.0}}
    \put(56,44){\scalebox{1.1}{0.5}}
    \put(56,34){\scalebox{1.1}{0.0}}
    \put(61,31.5){\scalebox{1.1}{0}}
    \put(67,31.5){\scalebox{1.1}{1}}
    \put(73,31.5){\scalebox{1.1}{2}}
    \put(79,31.5){\scalebox{1.1}{3}} 
    \put(85,31.5){\scalebox{1.1}{4}}
    \put(91,31.5){\scalebox{1.1}{5}}
    \put(50,32){\rotatebox{90}{\scalebox{1.1}{$-\langle w\rangle/(-\langle w\rangle)_\mathrm{max}$}}}
\end{overpic}
\caption{(a) Thermodynamic efficiency $\eta_{th}$ and (b) power output $P$ as a function of the driving time $\tau_{\mathrm{dri}}$ with $\mathcal{Q}(\rho_{12})=0, 0.031, 0.116$. The inset in (a) shows the nonadiabatic factor $\phi$ as a function of the driving time $\tau_{\mathrm{dri}}$. The parameters here are $\beta_{c}=0.6$, $\beta_{h}=0.3$, $\omega_{h}=3.8$, and $\omega_{c}=2$. For the selected parameters, the quantum discord values of $ \mathcal{Q}(\rho_{12})=0, 0.031,  
 0.116$ correspond to the interaction strengths of $ \xi=0, 1.5, 3$, respectively.}
    \label{pt}
\end{figure}

The nonadiabatic factor (\ref{phi1}) is dependent on the driving time $\tau_\mathrm{dri}$, and thus the efficiency $\eta_{th}$ is a function of both the quantum discord $\mathcal{Q}(\rho_{12})$ and the driving time $\tau_\mathrm{dri}$. When the driven stroke speeds downward, the nonadiabatic factor decreases, although not monotonically, and it approaches the lower bound of $\phi=1$ for large values of $\tau_\mathrm{dri}$ [see the inset of Fig. \ref{pt}(a)]. Consequently, the efficiency increases very quickly at first and then levels off, becoming asymptotic to the Otto efficiency [see Fig. \ref{pt}(a)]. In addition to the efficiency, the power output $P=-\langle {w}\rangle/\tau_{cyc}$ represents another important measure of the machine’s performance. The power output as a function of $\tau_\mathrm{dri}$ shows similar behavior to the efficiency [see Fig. \ref{pt}(b)].  
{The average work output $-\langle w\rangle$ (\ref{aw}) is dependent on the driving time $\tau_{\mathrm{dri}}$ because of the time-dependent nonadiabatic factor $\phi=\phi(\tau_{\mathrm{dri}})$, and it reaches its maximum value $(-\langle w\rangle)_{\mathrm{max}}$ when $\phi=1$. Therefore, for different values of the quantum discord, the values of the dimensionless work $-\langle w\rangle/(-\langle w\rangle)_{\mathrm{max}}$ obtained as a function of $\tau_\mathrm{dri}$ fall on top of each other at the point $\phi(\tau_{\mathrm{dri}})=1$, as illustrated in the inset of Fig. \ref{pt}(b).}
{We also observe that, when using the given parameters $\beta_{h,c}$ and $\omega_{c,h}$, the driving time at the maximum work is independent of the interaction strength and thus remains fixed, but the driving time at the maximum power output changes if we change the interaction strength associated with the quantum discord}. The power initially increases rapidly to approach the maximum value and then decreases nonmonotonically. In physical terms, when the driving time $\tau_\mathrm{dri}$ increases, the frictional work $\langle {w}_{\mathrm{fri}}\rangle$ drops rapidly, and then decreases very slowly to zero.

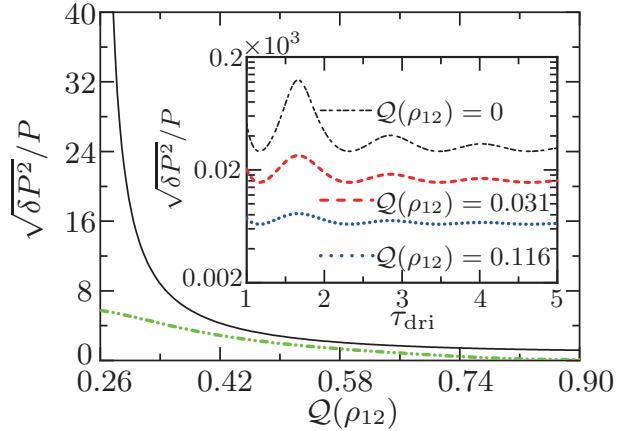
\begin{figure}
    \centering
\begin{overpic}[width=8cm]{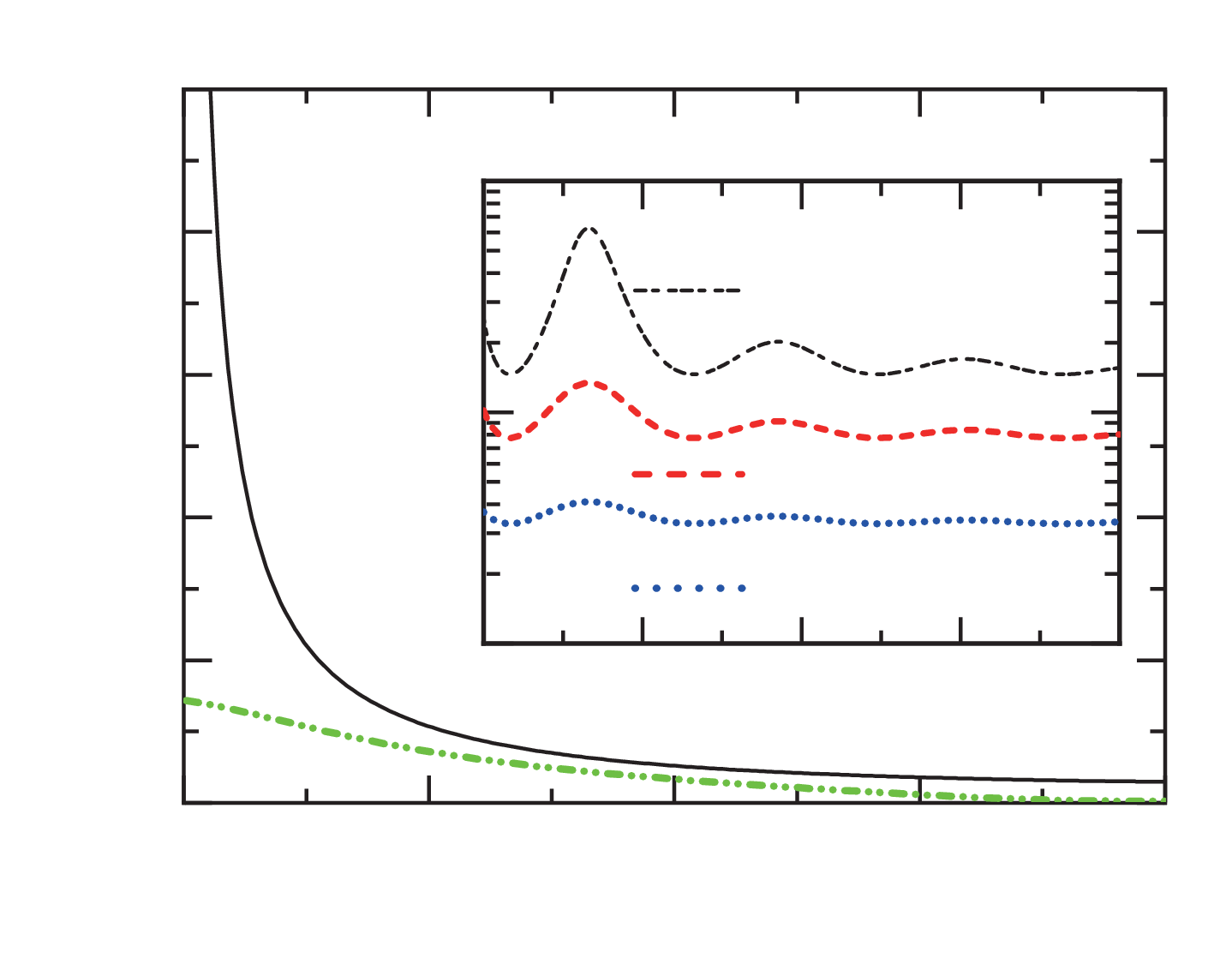}
\put(9,69){\scalebox{1.3}{40}}
    \put(9,57.5){\scalebox{1.3}{32}}
    \put(9,46){\scalebox{1.3}{24}}
    \put(9,34.5){\scalebox{1.3}{16}}
    \put(11,23){\scalebox{1.3}{8}}
    \put(11,12.5){\scalebox{1.3}{0}}
     \put(11,8){\scalebox{1.3}{0.26}}
    \put(31,8){\scalebox{1.3}{0.42}}
    \put(51,8){\scalebox{1.3}{0.58}}
    \put(71,8){\scalebox{1.3}{0.74}}
    \put(91,8){\scalebox{1.3}{0.90}}
     \put(63,19){\scalebox{1.3}{$\tau_\mathrm{dri}$}}
     \put(49,3){\scalebox{1.3}{$\mathcal{Q}(\rho_{12})$}}
     \put(33,62){\scalebox{1.1}{0.2}}
    \put(30.5,43.5){\scalebox{1.1}{0.02}}
    \put(28.5,25.5){\scalebox{1.1}{0.002}}
    \put(24,38){\rotatebox{90}{\scalebox{1.1}{$\sqrt{\delta P^2}/P$}}}
    \put(0,33){\rotatebox{90}{\scalebox{1.3}{$\sqrt{\delta P^2}/P$}}}
    \put(38,22){\scalebox{1.1}{1}}
     \put(51,22){\scalebox{1.1}{2}}
    \put(64,22){\scalebox{1.1}{3}}
    \put(77,22){\scalebox{1.1}{4}}
    \put(89.7,22){\scalebox{1.1}{5}}
    \put(38.5,64){\scalebox{1.1}{$\times 10^3$}}
    \put(60.5,53){\scalebox{1.1}{$\mathcal{Q}(\rho_{12})=0$}}
    \put(60.5,38){\scalebox{1.1}{$\mathcal{Q}(\rho_{12})=0.031$}}
    \put(60.5,29){\scalebox{1.1}{$\mathcal{Q}(\rho_{12})=0.116$}}
    \end{overpic}
    \caption{Coefficient of variation of power (black solid line) as a function of the quantum discord for $\tau_\mathrm{dri}=0.8$ compared with its lower bound $\mathrm{csch}[f(\langle \sigma\rangle)]$ (dot-dot-dashed line), where $f(x)$ is the inverse function of $x\tanh(x)$ and $x=\langle \sigma\rangle$ denotes the average entropy production in each engine cycle. The parameters are the same as those used in Fig. \ref{wqeta}(b). The inset, in which the ordinate axis is spaced logarithmically, shows the coefficient of variation of power as a function of the driving time for different quantum discord values. The other parameters are the same as those used in Fig. \ref{pt}.}
    \label{rp}
\end{figure}
We also observe from Figs. \ref{pt}(a) and \ref{pt}(b) that both the power output and the efficiency are enhanced by the quantum discord. This enhancement behavior is caused by the dependences of the excitation number $\langle n_{h}^{ss}\rangle$ in Eqs. (\ref{aw}) and (\ref{eff}) on the quantum discord $\mathcal{Q}(\rho_{12})$. Physically, the quantum discord, as a type of quantum resource, contributes to the mean work extracted and thus enhances the machine’s performance by increasing its power output and efficiency. Furthermore, we note that the oscillations of the curves for the efficiency and the power diminish with increasing quantum discord.
The final observation from Fig. \ref{pt} is
that the quantum entanglement is vanishing at the fixed quantum discord $\mathcal{Q}(\rho_{12})=0.031$, and the engine is thus fueled by the reservoirs in separable states. The presence of the quantum discord enhances the machine’s performance by increasing its efficiency and power and also improves its stability by reducing the relative fluctuations in power (see also the inset of Fig. \ref{rp}), even in the case where the reservoirs driving the quantum heat engines are in separable states, i.e., where the quantum entanglement is vanishing.

For quantum heat engines where the heat and the work are stochastic, the fluctuations in the power must be considered because they are associated with the machine’s stability. We consider the
coefficient of variation of power $\sqrt{\delta P^2}/ P$, which is equal to
the square root of the relative work fluctuations, given by $-\sqrt{\delta {w}^2}/\langle {w}\rangle.$ This dimensionless coefficient is shown as a function of the quantum discord in Fig. \ref{rp}, with the same coefficient being shown as a function of the driving time $\tau_{\mathrm{dri}}$ in the inset. The coefficient of variation of power decreases monotonically and its oscillations are damped as the quantum discord increases, thereby illustrating that the quantum discord can be used to damp the oscillations and the fluctuations in the power output, and thereby can improve the machine’s stability.

The total stochastic entropy production $\sigma$
is distributed according to the probability distribution $p(\sigma)$, where $\sigma (q_h,w)=(\beta_c-\beta_h^\mathrm{eff})q_h+\beta_c w$. These distributions for the working system, which involve time-reversal symmetry and satisfy the fluctuation theorems, can be expressed as $p(\sigma)=p_R(-\sigma)e^\sigma$ \cite{Crooks98}, with $p_R(-\sigma)$ being the probability distribution of the time-reversed cycle [in the clockwise direction in Fig. \ref{model}(b)]. These theorems always imply the generalized
thermodynamic uncertainty relationship for the stochastic work of the following form: 
$\delta w^2/\langle w\rangle^2\ge\mathrm{csch}^2[f(\langle\sigma\rangle)]$ \cite{Tim19}, where $f(x)$ is the inverse function
of $x\tanh(x)$. Therefore, the coefficient of variation of power should satisfy the following relation: $\sqrt {\delta P^2}/P\ge \mathrm{csch}[f(\langle\sigma\rangle)],$ as shown in Fig. \ref{rp}, where the function $\mathrm{csch}[f(\langle\sigma\rangle)]$ is indicated by the green dot-dot-dashed line.

\section{Discussion and Conclusions}\label{Co}

{As an initial remark, we must emphasize that all these results were obtained by assuming that the times $\tau_h$ and $\tau_c$ spent on the two thermalization strokes were negligible. The actual time periods for these two processes may have some importance in attempts to determine the machine performance in two very fast isochores because the quantum coherence \cite{Bar17, Su21, ar22} cannot be erased during the partial thermalization process. However, for the engine model under consideration here, where the two isochoric processes proceed at slow speeds, the quantum coherence can be assumed to be trivially small and even to be vanishing. In such a case, we prove in Appendix \ref{apet} that the effects of the values of $\tau_h$ and $\tau_c$ on the machine’s performance can be neglected, as we did in the main text of the paper.   }

{A natural extension of our model allows us to discuss another quantum Otto engine model,
in which an infinite collection of boson modes and a
beam composed of correlated pairs of atoms flying sequentially
through the cavity act as the hot thermal and cold nonthermal reservoirs, respectively. Unlike the case in our previous
model, where only one atom is weakly coupled to the optical cavity in the hot isochore, in the cold isochoric stroke
for the present engine cycle, both atoms in a pair will pass through
the optical cavity. As shown in Appendix \ref{apea}, numerical calculations of the average values of the heat, work, efficiency, and power demonstrate that the results obtained in the main text here are reproduced. Based on such a model, we  find that the quantum discord (even beyond
quantum entanglement) in either the cold reservoir or the hot
reservoir is quite beneficial to the machine’s performance and stability,
and it may also change the mode of the machine.}

{As another important model extension, the machine under consideration can be translated into a machine model in which the two reservoirs are both thermal but where nonclassical correlations exist in the working substance. One typical example of such a machine is a quantum Otto engine that works using a two-atom system (\ref{ham}) that is driven alternately by the two thermal reservoirs with constant inverse temperatures $\beta_h$ and $\beta_c$. The quantum Otto engine model can be controlled by controlling either the interaction strength $\xi$ or the frequency $\omega$ (see Appendix \ref{apeq} for details). 
 In contrast
to our model, which is driven by the nonthermal reservoir, the discord associated with the nonclassical correlations that exist in the working medium may 
reduce the efficiency under certain regimes and increase the power fluctuations, because interactions between the two atoms result in quantum fluctuations and thus lead to an inevitable increase in entropy \cite{Hong20}}.

In summary, we have set up a quantum engine working with a single-mode radiation field inside a resonant optical cavity, which is driven alternately by a thermal reservoir and an out-of-equilibrium reservoir with nonclassical correlations. The former reservoir is composed of boson modes and the latter is realized by sending one of two correlated atoms to interact weakly with the optical cavity.  {In a specific regime involving two bath temperatures, the machine will work as a heat engine for any values of the quantum discord, but for some bath temperature selections, modulation of the quantum discord may cause a change in the operation mode of the machine.} We have investigated the performance parameters, which are characterized by both the power and the thermodynamic efficiency, and the machine stability as measured by the variations in the power, all of which are dependent on the quantum discord associated with the nonclassical correlations. We have demonstrated that the quantum discord enables the machine to work as a heat engine in the extended
regime, where the machine without the quantum discord may operate as either a refrigerator or a heater. Moreover, our quantum Otto engine, when driven by the nonthermal reservoir with the nonclassical correlations, has enabled us to obtain superior levels of both efficiency and power when compared with a counterpart engine with thermal reservoirs, coinciding with the results obtained from previous models that showed superior performance characteristics
for nonthermal reservoirs, such as squeezed reservoirs \cite{Kla17, ar22, Aba14}. Finally, we have demonstrated that the results presented here can be reproduced using another quantum Otto engine model, which is driven by a hot thermal reservoir and a cold nonthermal reservoir by coupling a pair of spin atoms to the optical cavity. 
Our theoretical results may be helpful when using quantum discord as a type of resource to enable the design of efficient machines from a longer-term perspective.

\textbf{Acknowledgements}
This work was supported by the National Natural Science
Foundation (Grants No. 11875034, No. 12175040, and No. 12161056). J.W.
  acknowledges financial support from the Major Program of Jiangxi Provincial
Natural Science Foundation(Grants No. 20224ACB201007). Y.M. acknowledges financial support from the State Key Programs of China under Grant No. 2017YFA0304204.  
\appendix
\section{Derivation of the nonadiabatic factor}\label{aped}
{To describe the system dynamics along a unitary driven stroke, we rewrite the Hamiltonian of the optical cavity in terms of the momentum and position operators $\hat{p}$ and $\hat{x}$ to give ${H}^{ca}(t)=\hat{p}^2/(2m)+m\omega^2(t)\hat{x}^2$, and we also introduce the Lagrangian $\hat{L}^{ca}(t)=\hat{p}^2/(2m)+m\omega^2(t)\hat{x}^2$ and the position-momentum correlation $\hat{D}^{ca}(t)=\omega(t)(\hat{x}\hat{p}+\hat{p}\hat{x})/2$.
Next, we introduce a vector \cite{Lee20} $\overrightarrow{\psi}(t)=(\langle {H}^{ca}(t)\rangle, \langle \hat{L}^{ca}(t)\rangle, \langle \hat{D}^{ca}(t)\rangle, \langle \hat{I}\rangle )^\mathrm{T}$, where the superscript $\mathrm{T}$ denotes the matrix transpose. The dynamics for the vector $\overrightarrow{\psi}(t)$ can be given by: 
 \begin{equation}\label{master1}
     \frac{d}{dt}\overrightarrow{\psi}(t)=\mathcal{R}(t)\overrightarrow{\psi}(t),
 \end{equation}
where
\begin{eqnarray}\label{mat}
    \mathcal{R}(t)=\omega(t)
    \begin{pmatrix}
    \frac{\dot{\omega}(t)}{\omega^2(t)}&-\frac{\dot{\omega}(t)}{\omega^2(t)}&0&0\\
    -\frac{\dot{\omega}(t)}{\omega^2(t)}&\frac{\dot{\omega}(t)}{\omega^2(t)}&-2&0\\
    0&2&\frac{\dot{\omega}(t)}{\omega^2(t)}&0\\
    0&0&0&0
    \end{pmatrix}.
\end{eqnarray}
}
{We assume that 
 the system is in the thermal state at the beginning of the unitary driven stroke, as stated in our engine cycle, where the thermalization of the system is complete at the end of a hot or cold isochoric stroke. Using Eq. (\ref{master1}), at the end of a unitary stroke with time duration $\tau_{dri}$ and with the driving protocol $\omega_{ch}(t)=\omega_{c}\omega_{h}\tau_\mathrm{dri}/[(\omega_{c}-\omega_{h})t+\omega_{h}\tau_\mathrm{dri}]$, the system energy can be obtained via simple algebra \cite{Fei22}: 
\begin{equation}\label{hth}
\langle  H^{ca}(\tau_\mathrm{dri})\rangle
=\left[1+\frac{1-\cosh\left(\sqrt{1-\zeta}\ln\frac{\omega_h}{\omega_c}\right)}{\zeta-1}\right]\frac{\omega_h}{\omega_c}
\langle{H}^{ca}(0)\rangle,
\end{equation}
where we have  used $\zeta\equiv[2\tau_{\mathrm{dri}}\omega_{c}\omega_{h}/(\omega_{h}-\omega_{c})]^2$. The first factor in Eq. (\ref{hth}) can be identified as 
  the nonadiabatic factor $\phi$ [ see Eq. (\ref{phi1})].
 Within the quench limit with $\zeta\rightarrow 0$, we find that 
$\phi=(\omega_h^2+\omega_c^2)/(2\omega_c\omega_h)$.
In the special case where 
  $\zeta>1$, the nonadiabatic factor (\ref{phi1}) can be re-expressed as: } 

\begin{eqnarray}\label{hth0}
\phi&=&1+\frac{1-\cosh(i\sqrt{\zeta-1}\ln\frac{\omega_h}{\omega_c})}{\zeta-1}\nonumber\\
&=&1+\frac{1-\cos(\sqrt{\zeta-1}\ln\frac{\omega_h}{\omega_c})}{\zeta-1}. \label{phi0}
\end{eqnarray}

{\section{Time durations along the two isochoric processes}\label{apet}}
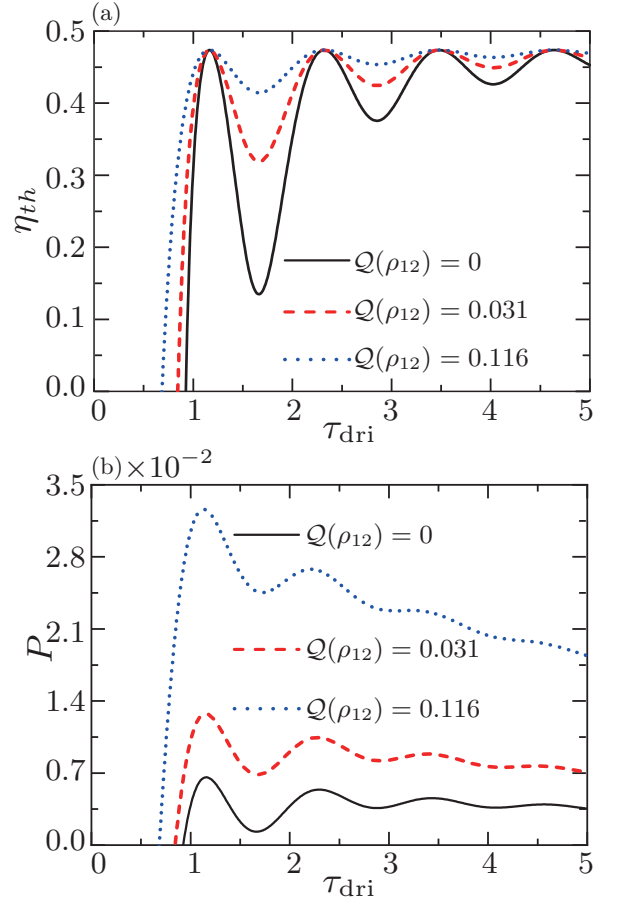
\begin{figure}
\begin{overpic}[width=8cm]{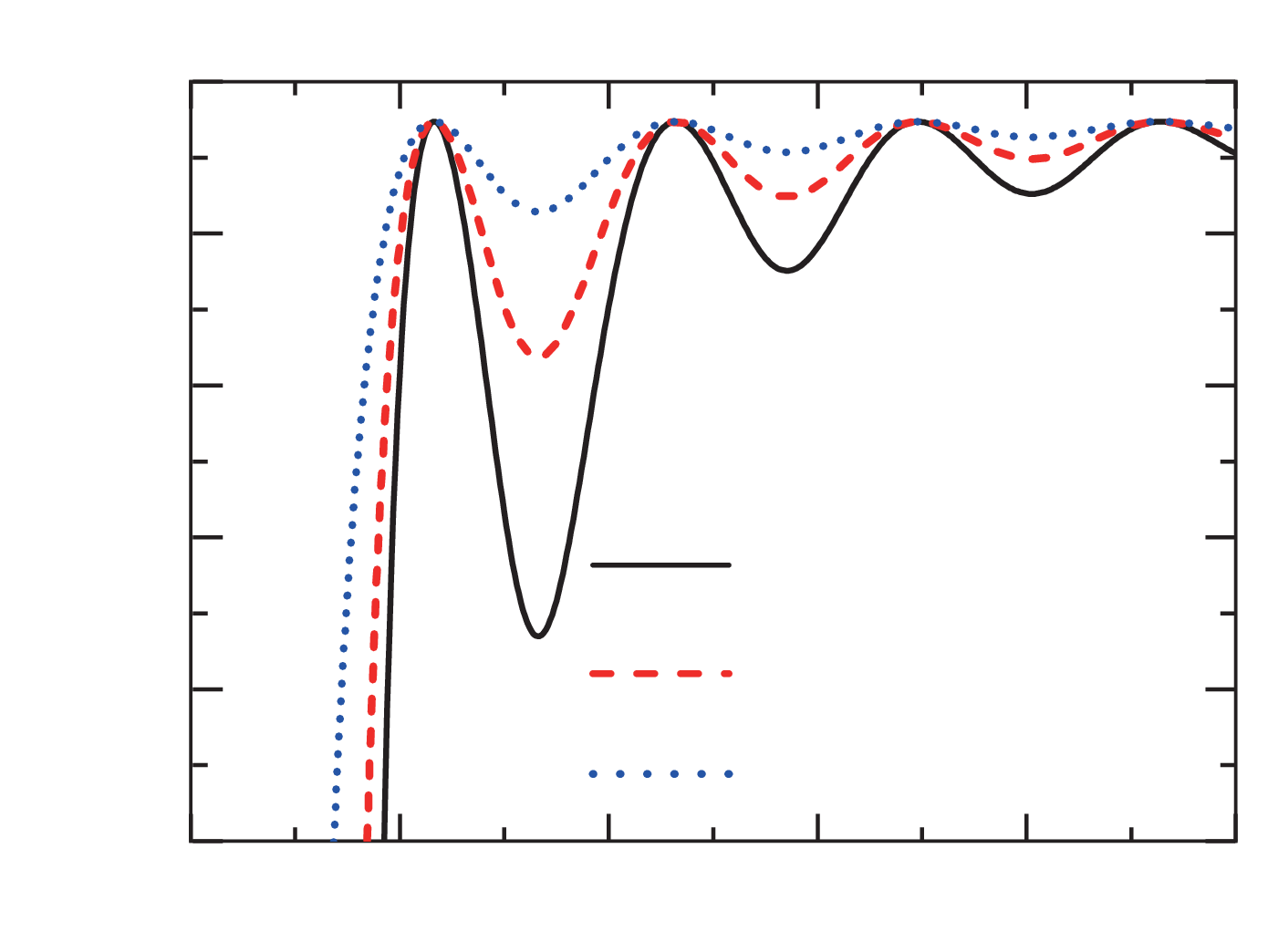}
    \put(14.5,70.5){(a)}
    \put(7,67){\scalebox{1.3}{0.5}}
    \put(7,55){\scalebox{1.3}{0.4}}
    \put(7,42.5){\scalebox{1.3}{0.3}}
    \put(7,31){\scalebox{1.3}{0.2}}
    \put(7,19.5){\scalebox{1.3}{0.1}}
    \put(7,8){\scalebox{1.3}{0.0}}
     \put(14.5,4){\scalebox{1.3}{0}}
    \put(30.5,4){\scalebox{1.3}{1}}
    \put(47,4){\scalebox{1.3}{2}}
    \put(63,4){\scalebox{1.3}{3}}
    \put(79.5,4){\scalebox{1.3}{4}}
    \put(95.5,4){\scalebox{1.3}{5}}
    \put(53,1){\scalebox{1.5}
    {$\tau_\mathrm{dri}$}}
    \put(58,29){\scalebox{1.1}{$\mathcal{Q}(\rho_{12})=0$}}
    \put(58,21){\scalebox{1.1}{$\mathcal{Q}(\rho_{12})=0.031$}}
    \put(58,13){\scalebox{1.1}{$\mathcal{Q}(\rho_{12})=0.116$}}
    \put(2,35){\rotatebox{90}{\scalebox{1.5}{$\eta_{th}$}}}
\end{overpic}
\begin{overpic}[width=8cm]{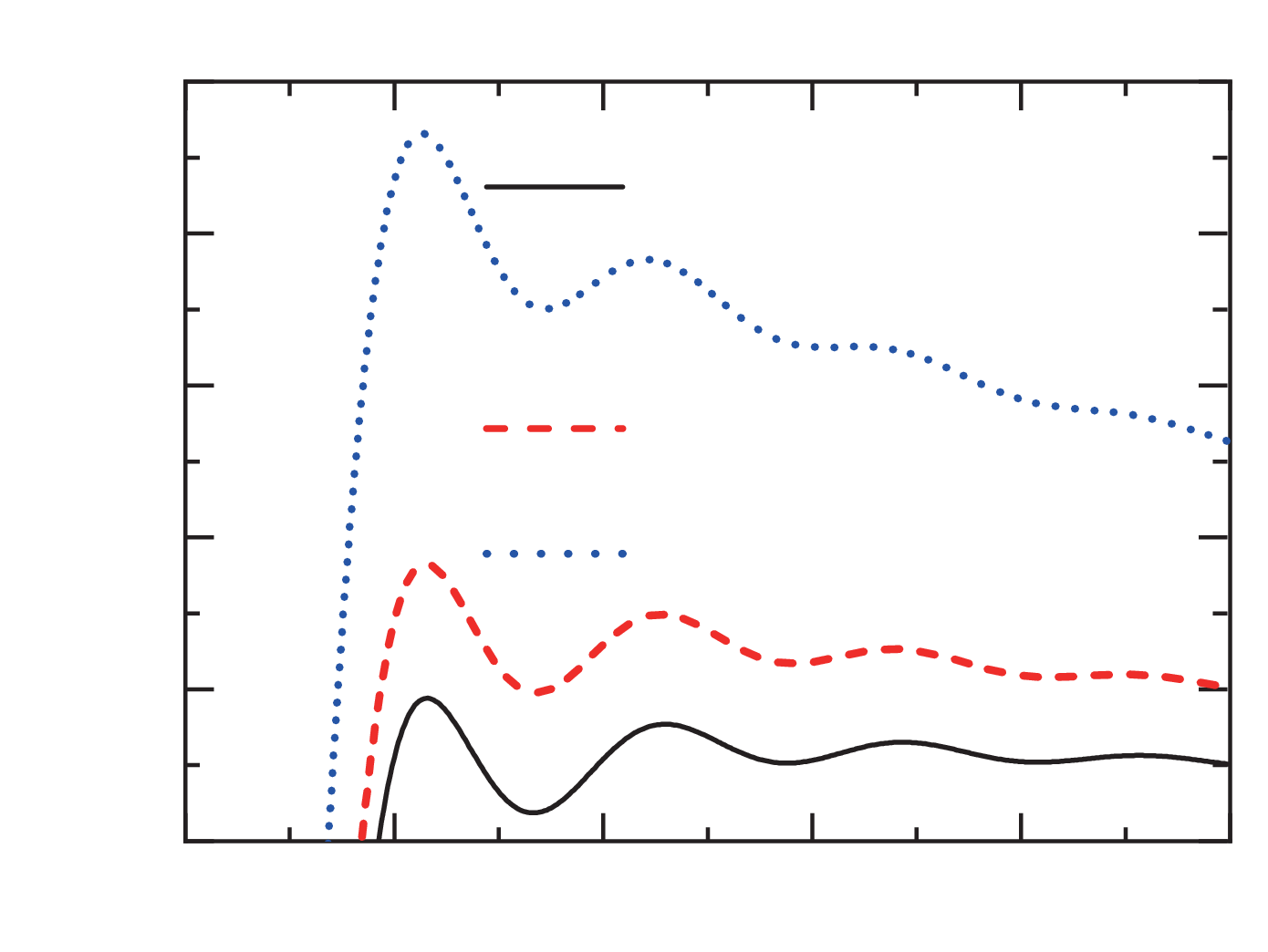}
    \put(14.5,70.5){(b)}
    \put(4,39){\rotatebox{90}{\scalebox{1.5}{$P$}}}
    \put(7,67){\scalebox{1.3}{3.5}}
    \put(7,55){\scalebox{1.3}{2.8}}
    \put(7,42.5){\scalebox{1.3}{2.1}}
    \put(7,31){\scalebox{1.3}{1.4}}
    \put(7,19.5){\scalebox{1.3}{0.7}}
    \put(7,8){\scalebox{1.3}{0.0}}
     \put(14.5,4){\scalebox{1.3}{0}}
    \put(30.5,4){\scalebox{1.3}{1}}
    \put(47,4){\scalebox{1.3}{2}}
    \put(63,4){\scalebox{1.3}{3}}
    \put(79.5,4){\scalebox{1.3}{4}}
    \put(95.5,4){\scalebox{1.3}{5}}
    \put(53,1){\scalebox{1.5}
    {$\tau_\mathrm{dri}$}}
    \put(20,70){\scalebox{1.3}{$\times 10^{-2}$}}
    \put(50,59){\scalebox{1.1}{$\mathcal{Q}(\rho_{12})=0$}}
    \put(50,40){\scalebox{1.1}{$\mathcal{Q}(\rho_{12})=0.031$}}
    \put(50,30){\scalebox{1.1}{$\mathcal{Q}(\rho_{12})=0.116$}}
\end{overpic}
\caption{{ (a) Thermodynamic efficiency and (b) power output as functions of driving time $\tau_\mathrm{dri}$ with different quantum discord values. The parameters used here are $\Gamma_c=\Gamma_h=1$ and $\tau_h=\tau_c=4$. The other parameters are same as those used in Fig. \ref{pt}.}}\label{ptth}   
\end{figure}
{Here, we analyze the time required to complete the hot and cold isochoric strokes, with reservoir temperatures of $\beta_\alpha$ ($\alpha=h,c$). To proceed, we consider only the transitions between two adjacent energy levels. Under this condition, 
 the dynamics for the probability of occupying a state $|n\rangle (n=0,1,2,...)$ are obtained by using Eq. (\ref{master}) \cite{Sch01}, where
\begin{eqnarray}\label{pmaster}
    \frac{dp_n}{dt}=k_u^\alpha  n p_{n-1}-[k_u^\alpha (n+1)+k_d^\alpha  n]p_n+k_d^\alpha (n+1)p_{n+1},\nonumber\\
\end{eqnarray}
where $k_u=r_1^\alpha(\gamma\tau)$ and $ k_d=r_2^\alpha(\gamma\tau)$. Here, $r_1^{\alpha}/r_2^{\alpha}$ satisfies the detailed balance requirements, which are $r_1^{h}/r_2^{h}=\mathrm{exp}(-\beta_{h}^\mathrm{eff}\omega_h)$ for the hot isochoric stroke and $r_1^{c}/r_2^{c}=\mathrm{exp}(-\beta_{c}\omega_c)$ for the cold isochoric process. Then, when using the mean photon number $\langle n \rangle =\underset{{n=0}}{\overset{\infty}{\sum}}(n+1/2)p_n=\langle \tilde{n}\rangle+1/2$ with $\langle \tilde{n}\rangle\equiv\underset{{n=0}}{\overset{\infty}{\sum}}np_n$, it follows that the evolution of the mean photon number over time is given by: 
\begin{eqnarray}\label{mmaster}
    \frac{d\langle n \rangle }{dt}&=&\frac{d\langle \tilde{n} \rangle}{dt}\nonumber\\
    &=&k_u^\alpha \underset{n'}{\sum} (n{'}+1)^2 p_{n'}-\underset{n}{\sum}[k_u^\alpha n (n+1)\nonumber\\
    &+&k_d^\alpha n^2]p_{n}
    +k_d^\alpha \underset{n''}{\sum}(n''-1)n''p_{n''},
\end{eqnarray}
 where we have used $n'=n-1$ and $n''=n+1$. The sums that contain the square of the summation index $n$ cancel, leading to:  
 \begin{equation}\label{mmaster2}
     \frac{d\langle \tilde{n}\rangle}{dt}=-(k_d^\alpha-k_u^\alpha)\langle \tilde{n}\rangle +k_u^\alpha,
 \end{equation}
{
the solution of which is given by 
 \begin{eqnarray} \label{nt}
    \langle \tilde{n}(t)\rangle=e^{-(k_d^\alpha-k_u^\alpha)t}\langle \tilde{n}(0)\rangle+\frac{k_u^\alpha}{k_d^\alpha-k_u^\alpha}[1-e^{-(k_d^\alpha-k_u^\alpha)t}].\nonumber\\
 \end{eqnarray}
 }
The mean photon numbers at the ends of the hot isochore $B\rightarrow C$ and the cold isochore $D\rightarrow A$ can then be obtained by using Eq. (\ref{nt}), where: 
 \begin{equation}\label{mh}
     \langle n_C\rangle =e^{-\Gamma_h\tau_h}(\langle n_B\rangle -1/2)+n_h^{qd}(1-e^{-\Gamma_h\tau_h})+1/2,
 \end{equation}
 and 
 \begin{equation}\label{mc}
     \langle n_A\rangle =e^{-\Gamma_c\tau_c}(\langle n_D\rangle-1) +n_c^{th}(1-e^{-\Gamma_c\tau_c})+1/2 ,
 \end{equation}
 respectively, 
 and we have used $\Gamma_\alpha=k_d^\alpha-k_u^\alpha$, $n_h^{qd}=k_u^h/(k_d^h-k_u^h)=[\exp(\beta_h^\mathrm{eff}\omega_h)-1]^{-1}$, and $n_c^{th}=k_u^c/(k_d^c-k_u^c)=[\exp(\beta_h^\mathrm{eff}\omega_h)-1]^{-1}$. Here, $\Gamma_\alpha$ denotes the heat conductivity between
the system and the heat reservoir at a temperature $\beta_\alpha (\alpha=c, h)$. Within the long time limit ($\tau_{h,c}\gg1$), the system reaches the steady state (i.e., thermal equilibrium) at the instant $C$ ($A$) within a single cycle [as illustrated in Fig. \ref{model}(b], yielding $\langle n_c^{eq}\rangle=n_c^{th}+\frac{1}2$ and $\langle n_h^{ss}\rangle=n_h^{qd}+\frac{1}2$.}

{The mean populations at the initial and final states of two unitary driven strokes satisfy the following relations \cite{Aba14}:
\begin{equation}\label{nB}
  \langle n_B\rangle=\phi\langle n_A\rangle, 
  \langle n_D\rangle=\phi\langle n_C\rangle,
\end{equation}
where $\phi$ was defined in Eq. (\ref{phi1}).
Through consideration of Eqs. (\ref{mh})$-$(\ref{nB}), the mean photon numbers at the respective ends of the cold and hot isochores, described by Eqs. (\ref{mh}) and (\ref{mc}), respectively, become  
\begin{equation}
    \langle n_A\rangle=\frac{\langle n_c^{eq}\rangle e^{\Gamma_h\tau_h}(e^{\Gamma_c\tau_c}-1)+\langle n_h^{ss}\rangle\phi(e^{\Gamma_h\tau_h}-1)}{(e^{\Gamma_c\tau_c+\Gamma_h\tau_h}-\phi^2)}, \label{na}
\end{equation}
\begin{equation}
 \langle n_C\rangle 
  =\frac{\langle n_h^{ss}\rangle e^{\Gamma_c\tau_c}(e^{\Gamma_h\tau_h}-1)+\langle n_c^{eq}\rangle\phi(e^{\Gamma_c\tau_c}-1)}{(e^{\Gamma_c\tau_c+\Gamma_h\tau_h}-\phi^2)}.
\end{equation}
The average population $\langle n_A\rangle$ ($\langle n_C\rangle$), as an exponentially decreasing (increasing) function of $\tau_c$ and $\tau_h$, approaches the specific limit $\langle n_c^{eq}\rangle$ ($\langle n_h^{ss}\rangle$).} 
{By replacing $\langle n_c^{eq}\rangle$ and $\langle n_h^{ss}\rangle$ with $\langle n_A\rangle$ and $\langle n_C\rangle$ in Eqs. (\ref{aw}), (\ref{var}), and (\ref{eff}), the output work, the fluctuation of work, and the thermodynamic efficiency become} 
\begin{eqnarray}\label{aw1}
\langle{w}\rangle
=\omega_{h}(\phi\langle n_A\rangle-\langle n_C\rangle)+\omega_{c}(\phi\langle n_C\rangle-\langle n_A\rangle), 
\end{eqnarray}
\begin{eqnarray}\label{var1}
\delta w^{2}&=&\omega_{h}^2[-\frac{1}{2}+(2\phi^2-1)\langle n_A\rangle^2+\langle n_C\rangle^2]\nonumber\\
&+&\omega_{c}^2[-\frac{1}{2}+\langle n_A\rangle^2+(2\phi^2-1)\langle n_C\rangle^2]\\\nonumber
&+&\omega_{h}\omega_{c}\phi(1-2\langle n_A\rangle^2-2\langle n_C\rangle^2),  
\end{eqnarray} 
\begin{equation}\label{eff1}
\eta_{th}=1-\frac{\omega_c}{\omega_h}-\frac{\omega_{c}}{\omega_{h}}\frac{(\phi-1)(\langle n_C\rangle+\langle n_A\rangle)}{\langle n_C\rangle-\phi\langle n_A\rangle},
\end{equation}
respectively. {Using Eqs. (\ref{aw1}), (\ref{var1}), and (\ref{eff1}), we can then numerically determine the power output [$P=-\langle w\rangle/(2\tau_{\mathrm{dri}}+\tau_c+\tau_h)$], the efficiency, and the relative power fluctuations. As an example, by setting $\gamma_h=\gamma_c=1$ and $\tau_c=\tau_h=4$, we plot the power output and the efficiency in Fig. \ref{ptth} as functions of the driving time $\tau_\mathrm{dri}$ for different values of the quantum discord. 
 The shapes of the efficiency and power output curves in Figs. \ref{ptth}(a) and \ref{ptth}(b) are similar to the corresponding results in Figs. \ref{pt}(a) and \ref{pt}(b), respectively. Because $\langle n_A\rangle$ ($\langle n_C\rangle$) increases very rapidly with increasing $\tau_c$ and $\tau_h$, it approaches $\langle n_c^{eq}\rangle (\langle n_h^{ss}\rangle)$ when the times $\tau_{c,h}$ are long enough. The long time required to complete a thermalization process gives the  results similar to those obtained using our model when assuming $\tau_{c,h}$ to be negligible, which means that selection of $\tau_c$ and $\tau_h$ such that they fell into a very specific range produced the same results, and thus our theory is applicable to our machine model, where the working medium is very close to the steady state at the end of each isochore. }

{\section{An Otto engine in which the optical cavity is coupled to the pair of interacting atoms along the cold isochore}\label{apea}}
\begin{figure*}
\begin{overpic}[width=8cm]{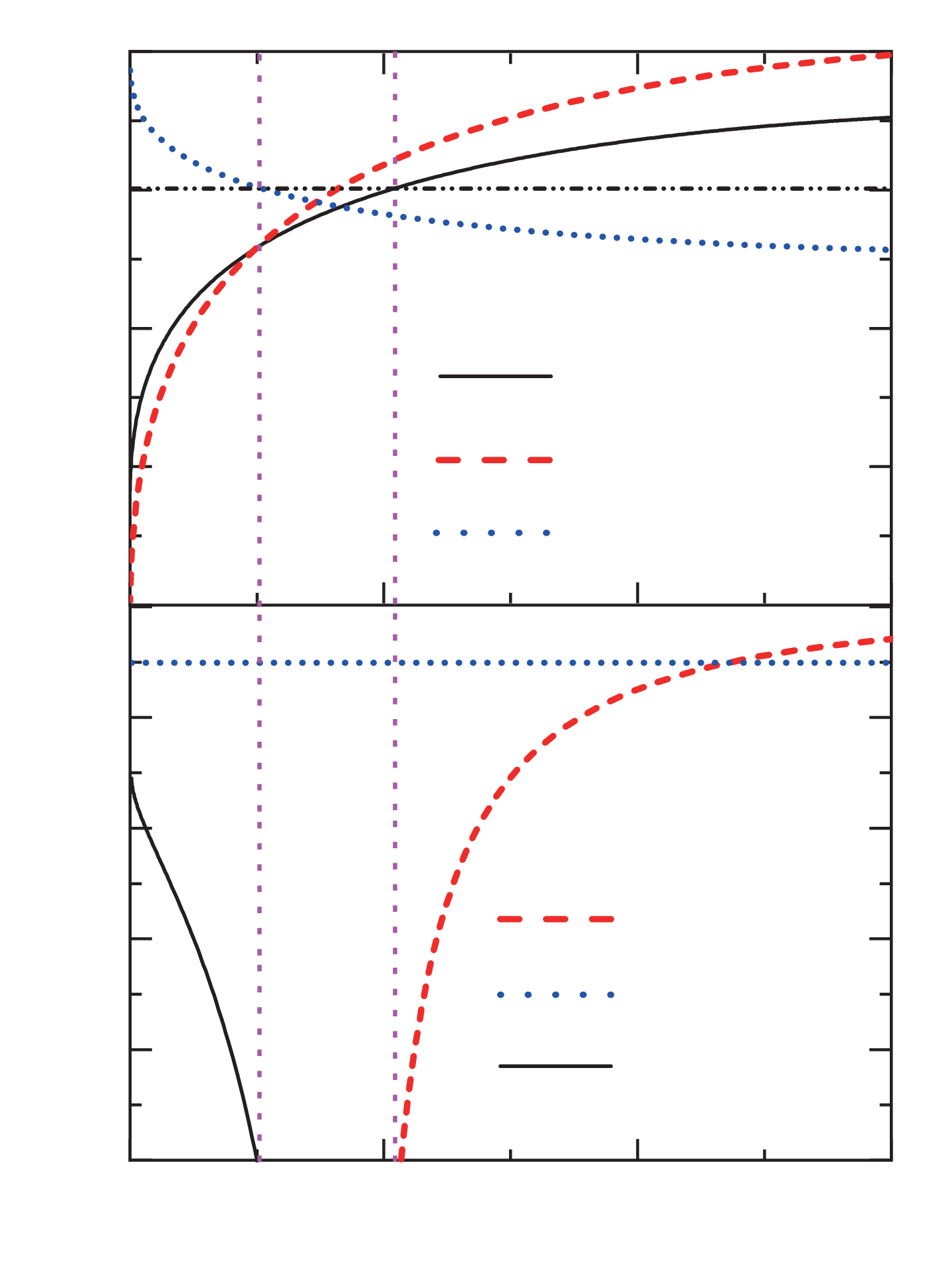}
    \put(9.5,97.5){(a)}
    \put(4.5,94){\scalebox{1.3}{0.5}}
    \put(4.5,84.5){\scalebox{1.3}{0.0}}
    \put(1.5,74){\scalebox{1.3}{$-0.5$}}
     \put(1.5,63.5){\scalebox{1.3}{$-1.0$}}
    \put(1.5,53.5){\scalebox{1.3}{$-1.5$}}
    \put(2.5,50.5){\scalebox{1.3}{0.55}}
    \put(2.5,43){\scalebox{1.3}{0.44}}
    \put(2.5,34.5){\scalebox{1.3}{0.33}}
     \put(2.5,26){\scalebox{1.3}{0.22}}
    \put(2.5,17.5){\scalebox{1.3}{0.11}}
     \put(2.5,9){\scalebox{1.3}{0.00}}
\put(43,70){\scalebox{1.3}{$-\langle w\rangle$}}
\put(43,63.5){\scalebox{1.3}{$\langle q_h \rangle$}}
\put(43,58){\scalebox{1.3}{$\langle q_c \rangle$}}
\put(48,28){\scalebox{1.3}{$\eta_{th}$ }}
\put(48,22){\scalebox{1.3}{$\eta_{C}$ }}
\put(48,16){\scalebox{1.3}{$\varepsilon$}}
\put(6.5,5.5){\scalebox{1.3}{0.0}}
\put(26,5.5){\scalebox{1.3}{0.3}}
\put(46,5.5){\scalebox{1.3}{0.6}}
\put(66,5.5){\scalebox{1.3}{0.9}}
\put(33,1){\scalebox{1.3}{$\mathcal{Q}(\rho_{12})$}}
\put(25,70){\rotatebox{90}{\scalebox{0.9}{Heater}}}
\put(25,70){\rotatebox{90}{\scalebox{0.9}{Heater}}}
\put(16,60.5){\rotatebox{90}{\scalebox{0.9}{refrigerator}}}
\put(16,60.5){\rotatebox{90}{\scalebox{0.9}{refrigerator}}}
\put(46,75){\scalebox{0.9}{Heat engine}}
\put(46,75){\scalebox{0.9}{Heat engine}}
\end{overpic}
\begin{overpic}[width=8cm]{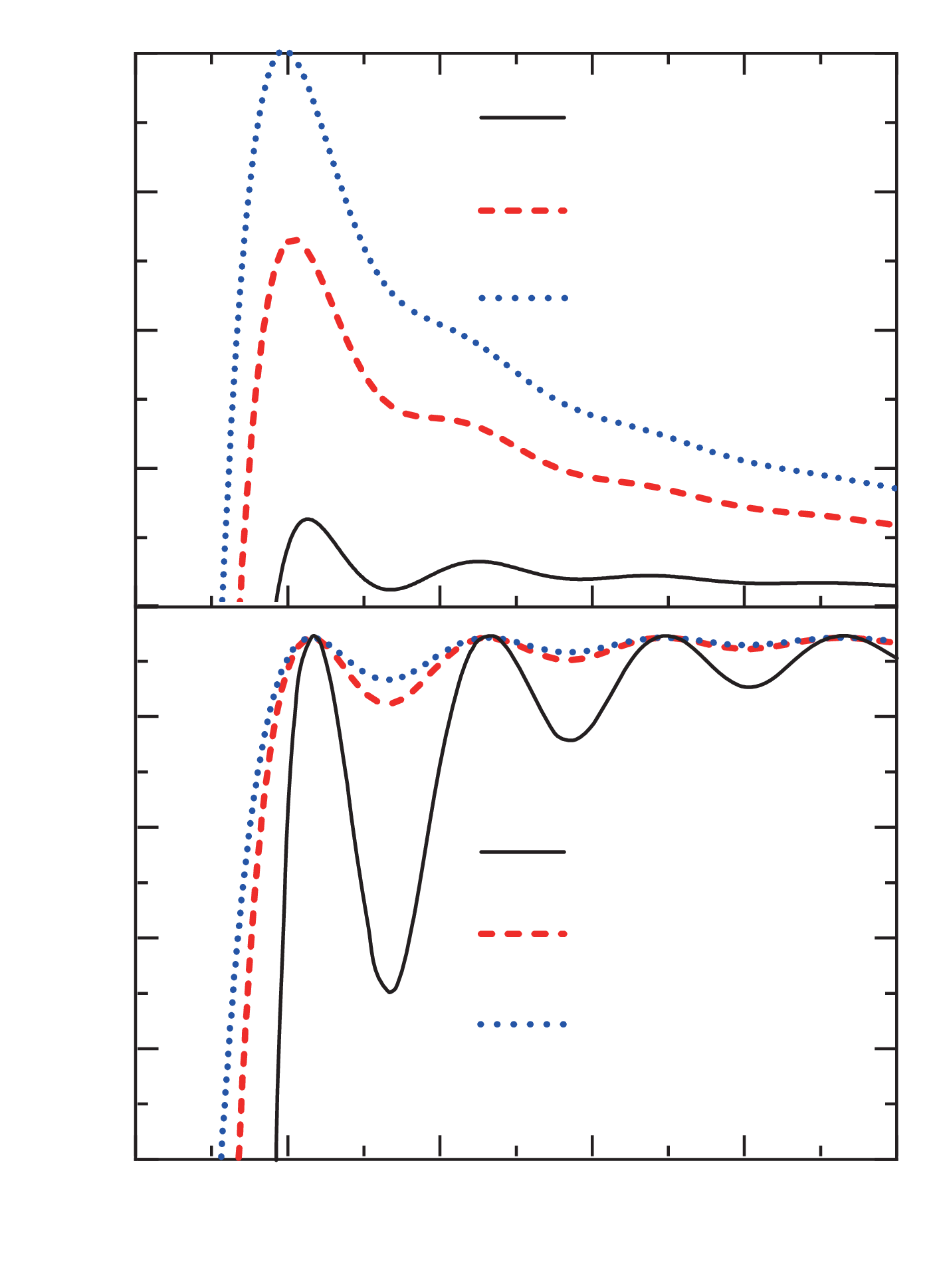}
    \put(9.5,97.5){(b)}
    \put(2.5,94){\scalebox{1.3}{0.20}}
    \put(2.5,84.5){\scalebox{1.3}{0.15}}
    \put(2.5,74){\scalebox{1.3}{0.10}}
     \put(2.5,63.5){\scalebox{1.3}{0.05}}
    \put(2.5,53.5){\scalebox{1.3}{0.00}}
    \put(4.5,50.5){\scalebox{1.3}{0.5}}
    \put(4.5,43){\scalebox{1.3}{0.4}}
    \put(4.5,34.5){\scalebox{1.3}{0.3}}
     \put(4.5,26){\scalebox{1.3}{0.2}}
    \put(4.5,17.5){\scalebox{1.3}{0.1}}
     \put(4.5,9){\scalebox{1.3}{0.0}}
      \put(44,90){\scalebox{1.1}{$\mathcal{Q}(\rho_{12})=0$}}
    \put(44,83){\scalebox{1.1}{$\mathcal{Q}(\rho_{12})=0.031$}}
    \put(44,76){\scalebox{1.1}{$\mathcal{Q}(\rho_{12})=0.116$}}
    \put(44,33){\scalebox{1.1}{$\mathcal{Q}(\rho_{12})=0$}}
    \put(44,26.5){\scalebox{1.1}{$\mathcal{Q}(\rho_{12})=0.031$}}
    \put(44,19.5){\scalebox{1.1}{$\mathcal{Q}(\rho_{12})=0.116$}}
    \put(1,70){\rotatebox{90}{\scalebox{1.5}{$P$}}}
    \put(1,30){\rotatebox{90}
    {\scalebox{1.5}{$\eta_{th}$}}}
     \put(10,6){\scalebox{1.3}{0}}
    \put(21.5,6){\scalebox{1.3}{1}}
    \put(33,6){\scalebox{1.3}{2}}
    \put(45,6){\scalebox{1.3}{3}}
    \put(57,6){\scalebox{1.3}{4}}
    \put(69,6){\scalebox{1.3}{5}}
    \put(37,1){\scalebox{1.5}{$\tau_\mathrm{dri}$}}
\end{overpic}
\caption{(a) Work output and heat values absorbed by the system (upper panel), and thermodynamic efficiency (lower panel) as
functions of the quantum discord for $\tau_\mathrm{dri}=0.8$ and $\omega_h=6$. (b) Output power (upper panel) and thermodynamic efficiency (lower panel) as functions of the driving time for different values of the quantum discord at $\omega_h=3.8$. The other parameters are the same as those used in Fig. \ref{rp}.}
    \label{pct}
\end{figure*}
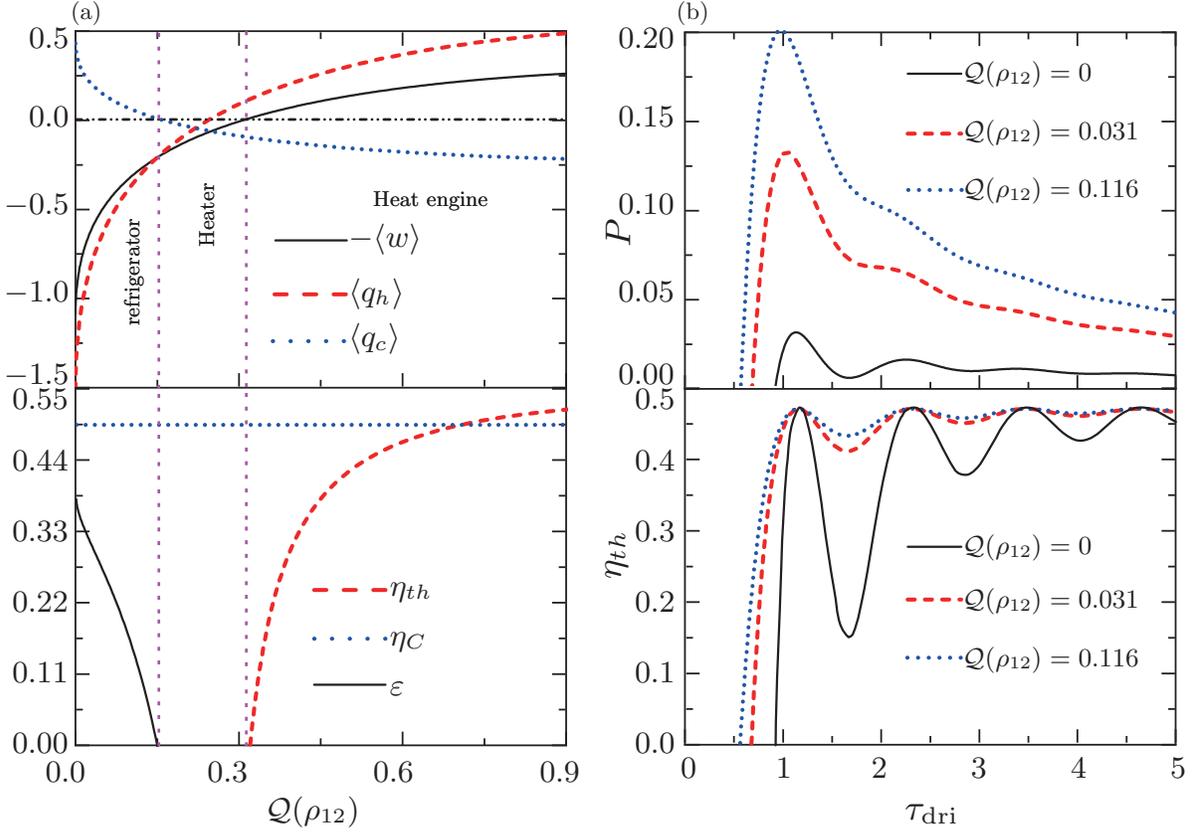
{Here, we consider an alternative quantum Otto engine in which the optical cavity interacts simultaneously with the pair of correlated atoms (\ref{ham}) of inverse temperature $\beta_c$ during the cold isochore. 
 The inverse temperatures of the two reservoirs are also still denoted by
$ \beta_h$ and $\beta_c(> \beta_h)$. In each cycle, the single-mode radiation field in the optical cavity acting as the working substance
reaches the steady (thermal) state at the end of the
cold (hot) isochoric stroke}.

To proceed, we define $r_1^{c}=\rho_e^{c}+\rho_d^{c}+\rho_{nd}^c$ and $r_2^{c}=\rho_g^{c}+\rho_d^{c}+\rho_{nd}^{c}$,
 with $\rho_{e}^{c}=\mathrm{exp}(-\beta_{c}\omega_{c})/[2\mathrm{cosh}(\beta_{c}\omega_{c})+2\mathrm{cosh}(\beta_{c}\xi)]$, $\rho_{g}^c=\mathrm{exp}(\beta_c\omega_c)/[2\mathrm{cosh}(\beta_c\omega_c)+2\mathrm{cosh}(\beta_c\xi)]$, $\rho_{d}^c=\mathrm{cosh}(\beta_c\xi)/[2\mathrm{cosh}(\beta_c\omega_c)+2\mathrm{cosh}(\beta_c\xi)]$, and $\rho_{nd}^c=-\mathrm{sinh}(\beta_c\xi)/[2\mathrm{cosh}(\beta_c\omega_c)+2\mathrm{cosh}(\beta_c\xi)]$. By replacing $r_1^{h}$ and $r_2^{h}$ with $r_1^{c}$ and $r_2^{c}$, respectively, in Eq. (\ref{master}), the asymptotic stationary solution is given by $\rho^{ss}_c=(e^{\beta_{c}^\mathrm{eff} \omega_{c}/2}-e^{-\beta_{c}^\mathrm{eff} \omega_{c}/2})e^{-\beta_{c}^\mathrm{eff}H^{ca}(\omega_c)}$, where $\beta_{c}^\mathrm{eff}$ denotes the effective
inverse temperature of the optical cavity.
 With the detailed balance condition,
the ratio of $r_1^{c}/r_2^{c}$ becomes $r_1^{c}/r_2^{c}=\mathrm{exp}(-\beta_{c}^\mathrm{eff}\omega_{c})$, which leads to \cite{Lutz09}
 \begin{eqnarray}\label{betac}
 {\beta_c^\mathrm{eff}}=\beta_{c}-\frac{1}{\omega_{c}}\mathrm{ln}\frac{1+e^{{\beta_c(\omega_c-\xi)}}}{{e^{\beta_c\omega_c}}+{e^{-\beta_c\xi}}},
 \end{eqnarray}
 which then produces the excitation number $\langle n_{c}^{ss}\rangle=[\exp({\beta_{c}^{\mathrm{eff}}}\omega_{c})-1]^{-1}$. While remaining in contact with the hot thermal reservoir, the working substance relaxes to the thermal state at the end of the hot isochoric stroke, and its
the excitation number is then given by $\langle n_h^{eq}\rangle=[\exp(\beta_h\omega_h)-1]^{-1}$. The important point here is that the nonadiabatic factor (\ref{phi1}) is only dependent on the system Hamiltonian and the driving time. Therefore, the work statistics ($\langle {w}\rangle$ and $\langle\delta {w}\rangle^2$) and the efficiency ($\eta_{th})$ are still given by Eqs. (\ref{aw}), (\ref{var}), and (\ref{eff}), respectively, while replacing $\langle n_{h}^{ss}\rangle $ and $\langle n_{c}^{eq}\rangle$ with $\langle n_{h}^{eq}\rangle$ and $\langle n_{c}^{ss}\rangle$, respectively. In an analogous manner to the efficiency of the heat engine, the coefficient of performance, which is defined as $\varepsilon=\langle q_c\rangle/\langle w\rangle$, is a performance measure for the refrigerator. The numerical results for the average work, the average heat values during the hot and cold isochores, the power output, and the efficiency (coefficient of performance) are presented in Fig. \ref{pct}.  

The operating mode of the thermal machine can be changed
 if the strength of the quantum discord varies; we find that, for a given set of control parameters, the quantum-discord-dependent operating mode may cause the machine to act as a refrigerator, a heater, or a heat engine, as shown in Fig. \ref{pct}(a). In addition, the machine can operate with
high efficiency, including at efficiencies greater than the Carnot efficiency, if the quantum discord is large enough, which is consistent with the
results presented in Fig. \ref{wqeta}.
 Figure \ref{pct}(b) shows that the power and efficiency are both improved as the quantum discord in the cold isochoric state is increased, which is consistent with the results in Fig. \ref{pt}. Our additional calculations of the power fluctuations, which are not plotted here, demonstrate that the variance of the power behaves in a manner corresponding to that shown in Fig. \ref{rp}. Therefore, for this machine model, the
quantum discord (even beyond the quantum entanglement) in either the cold or the hot heat reservoir is quite beneficial to machine performance and stability, and can even cause a change in the machine mode.

{\section{Quantum Otto engines based on the two-atom system with nonclassical correlations}\label{apeq}}
\begin{figure*}
  \centering
  \begin{overpic}[width=7cm]{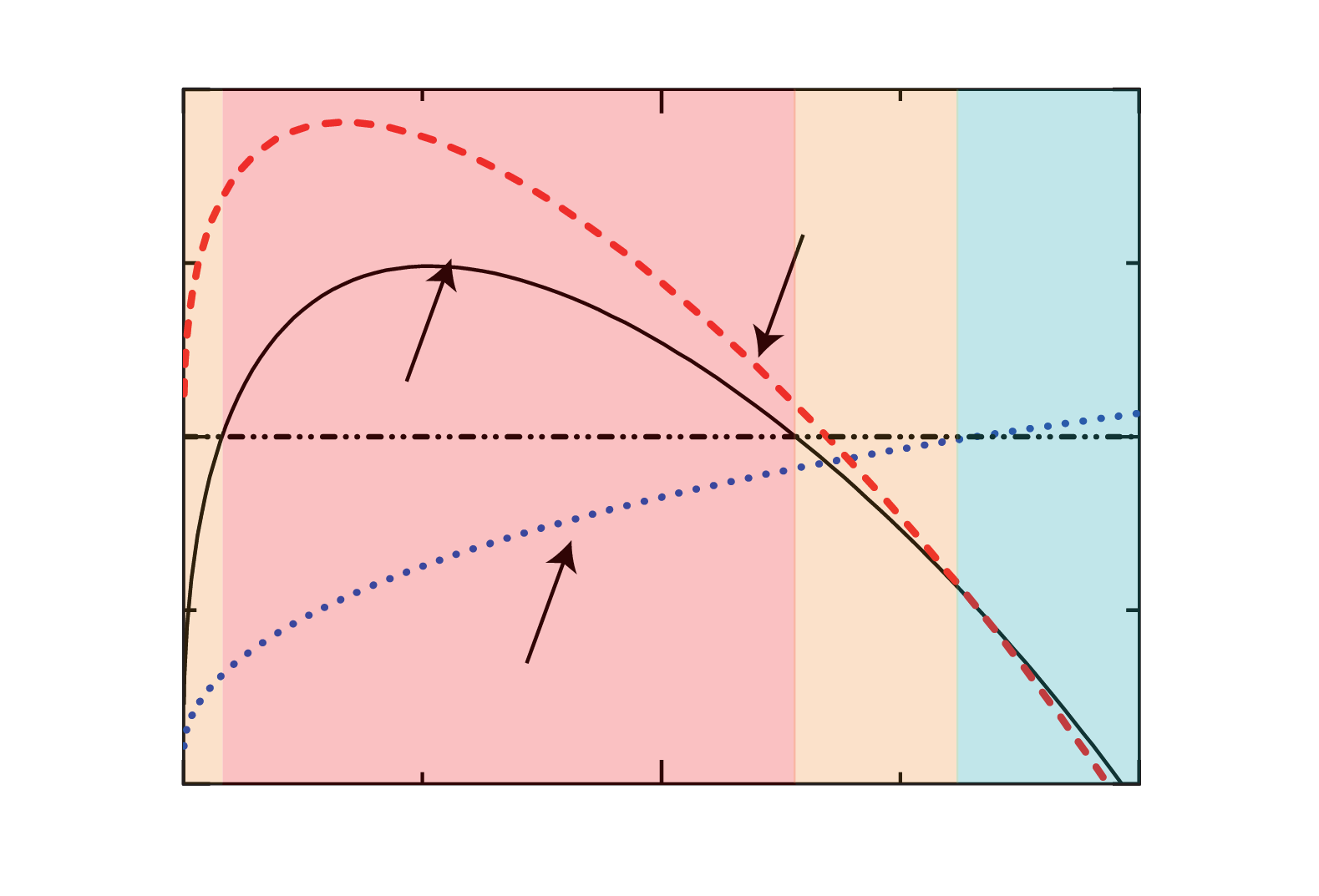}
\put(13.2,63.3){(a)}
\put(5,59){\scalebox{1.3}{3.5}}
\put(5,32.8){\scalebox{1.3}{0.0}}
\put(3,8){\scalebox{1.3}{-3.5}}
\put(11,4){\scalebox{1.3}{0.00}}
\put(46,4){\scalebox{1.3}{0.35}}
\put(80,4){\scalebox{1.3}{0.70}}
\put(42,-1){\scalebox{1.3}{$\mathcal{Q}(\rho_{12})$}}
\put(27,37){\scalebox{1.3}{$-\langle w\rangle$}}
\put(56,52){\scalebox{1.3}{$\langle q_h\rangle$}}
\put(35,12){\scalebox{1.3}{$\langle q_c\rangle$}}
\end{overpic}
\begin{overpic}[width=7cm]{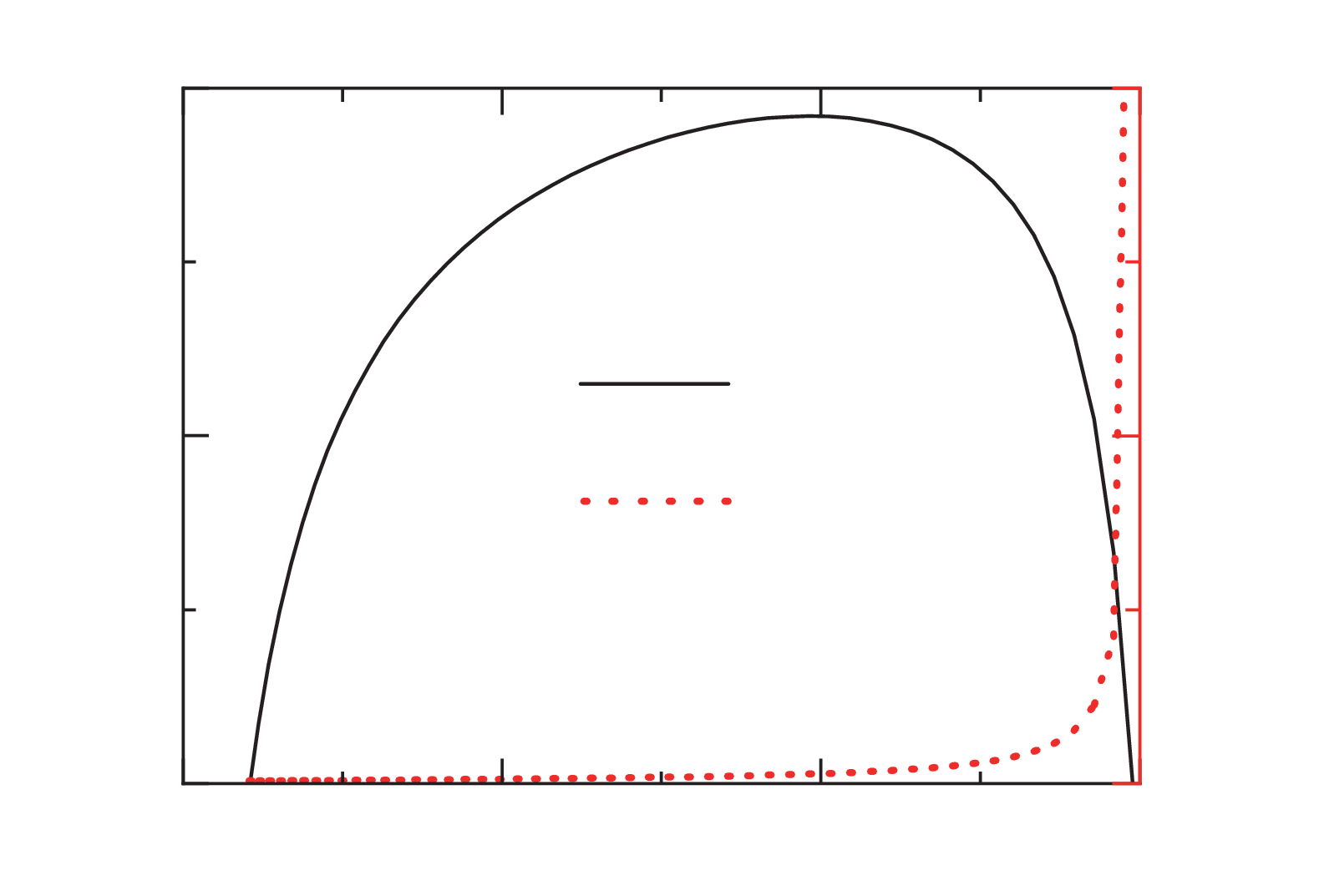}
\put(13.2,63.3){(b)}
\put(2,59){\scalebox{1.3}{0.65}}
\put(2,32.8){\scalebox{1.3}{0.35}}
\put(2,8){\scalebox{1.3}{0.05}}
\put(11,4){\scalebox{1.3}{0.00}}
\put(34,4){\scalebox{1.3}{0.15}}
\put(58,4){\scalebox{1.3}{0.30}}
\put(80,4){\scalebox{1.3}{0.45}}
\put(42,-1){\scalebox{1.3}{$\mathcal{Q}(\rho_{12})$}}
\put(57,38){\scalebox{1.3}{$\eta_{th}$}}
\put(57,28){\scalebox{1.1}{$\sqrt{\delta P^2}/P$}}
\put(87,59){\scalebox{1.3}{\textcolor{red}{400}}}
\put(87,32.8){\scalebox{1.3}{\textcolor{red}{200}}}
\put(87,8){\scalebox{1.3}{\textcolor{red}{0}}}
\put(96,44){\scalebox{1.1}{\rotatebox{270}{$\sqrt{\delta P^2}/P$}}}
\put(-2,31){\scalebox{1.3}{\rotatebox{90}{$\eta_{th}$}}}
\end{overpic}
 \begin{overpic}[width=7cm]{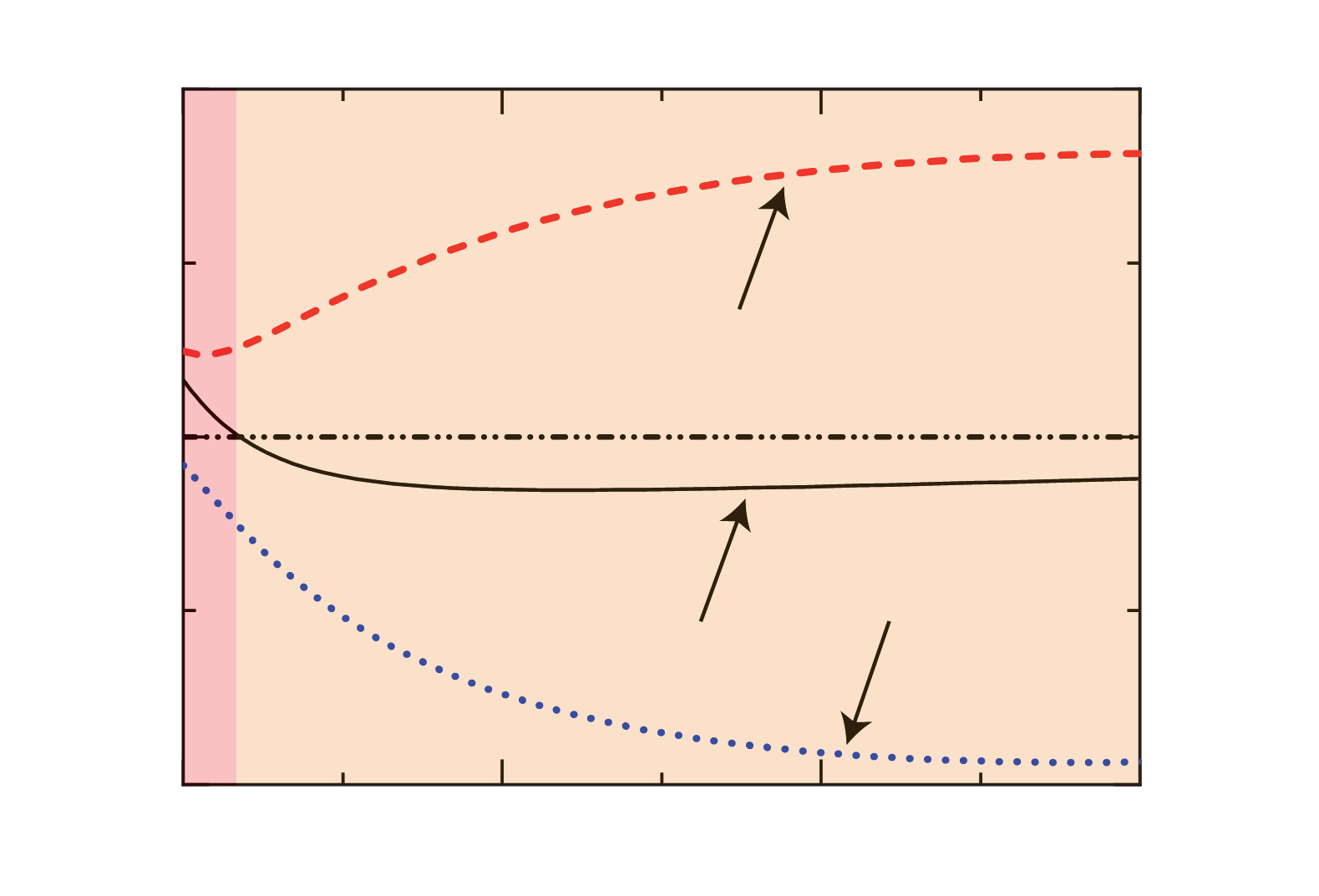}
\put(13.2,63.3){(c)}
\put(9,59){\scalebox{1.3}{6}}
\put(9,32.8){\scalebox{1.3}{0}}
\put(4.5,8){\scalebox{1.3}{$-6$}}
\put(11,4){\scalebox{1.3}{0.0}}
\put(34,4){\scalebox{1.3}{0.1}}
\put(58,4){\scalebox{1.3}{0.2}}
\put(80,4){\scalebox{1.3}{0.3}}
\put(42,-1){\scalebox{1.3}{$\mathcal{Q}(\rho_{12})$}}
\put(45,16){\scalebox{1.3}{$-\langle w\rangle$}}
\put(51,40){\scalebox{1.3}{$\langle q_h\rangle$}}
\put(62,23){\scalebox{1.3}{$\langle q_c\rangle$}}
\end{overpic}
\begin{overpic}[width=7cm]{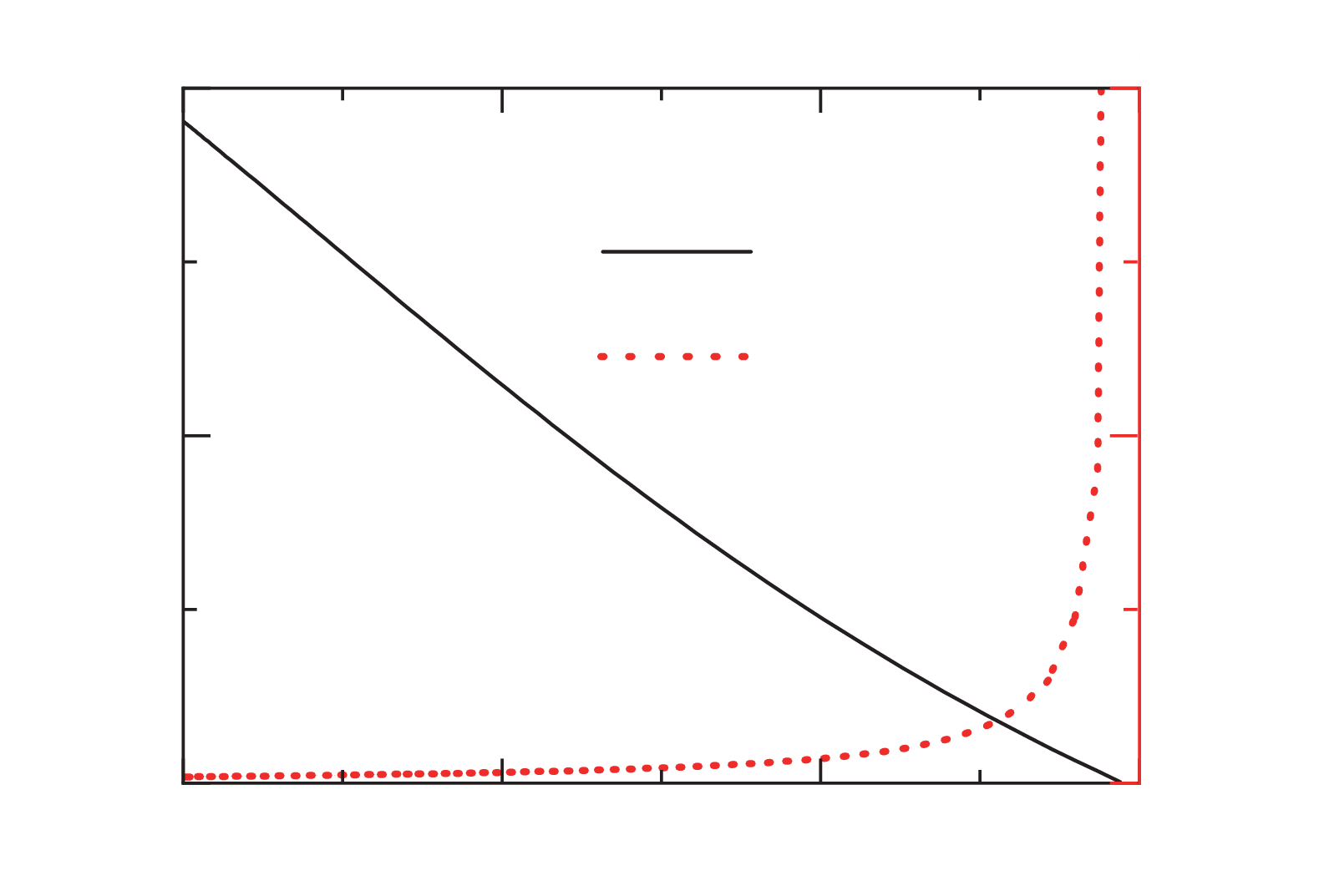}
\put(13.2,63.3){(d)}
\put(2,59){\scalebox{1.3}{0.70}}
\put(2,32.8){\scalebox{1.3}{0.35}}
\put(2,8){\scalebox{1.3}{0.00}}
\put(11,4){\scalebox{1.3}{0.000}}
\put(34,4){\scalebox{1.3}{0.006}}
\put(58,4){\scalebox{1.3}{0.012}}
\put(80,4){\scalebox{1.3}{0.018}}
\put(42,-1){\scalebox{1.3}{$\mathcal{Q}(\rho_{12})$}}
\put(58,48){\scalebox{1.3}{$\eta_{th}$}}
\put(57,39){\scalebox{1.1}{$\sqrt{\delta P^2}/P$}}
\put(87,59){\scalebox{1.3}{\textcolor{red}{400}}}
\put(87,32.8){\scalebox{1.3}{\textcolor{red}{200}}}
\put(87,8){\scalebox{1.3}{\textcolor{red}{0}}}
\put(96,44){\scalebox{1.1}{\rotatebox{270}{$\sqrt{\delta P^2}/P$}}}
\put(-2,31){\scalebox{1.3}{\rotatebox{90}{$\eta_{th}$}}}
\end{overpic}
\caption{(a and c) Numerical results for the work output and heat values absorbed by the system. (b and d) Thermodynamic efficiency (left axle) and coefficient of variation of power (right axle) as functions of the quantum discord. The parameters for (a) and (b) are $\beta_c=0.6,
\beta_h=0.1, \omega=2, \xi_c=4$. The parameters for (c) and (d) are $\beta_c=0.6,
\beta_h=0.1, \omega_c=2, \omega_h=6.$ The parameter $\tau_\mathrm{dri}=1$.}
\label{qubit}
\end{figure*}
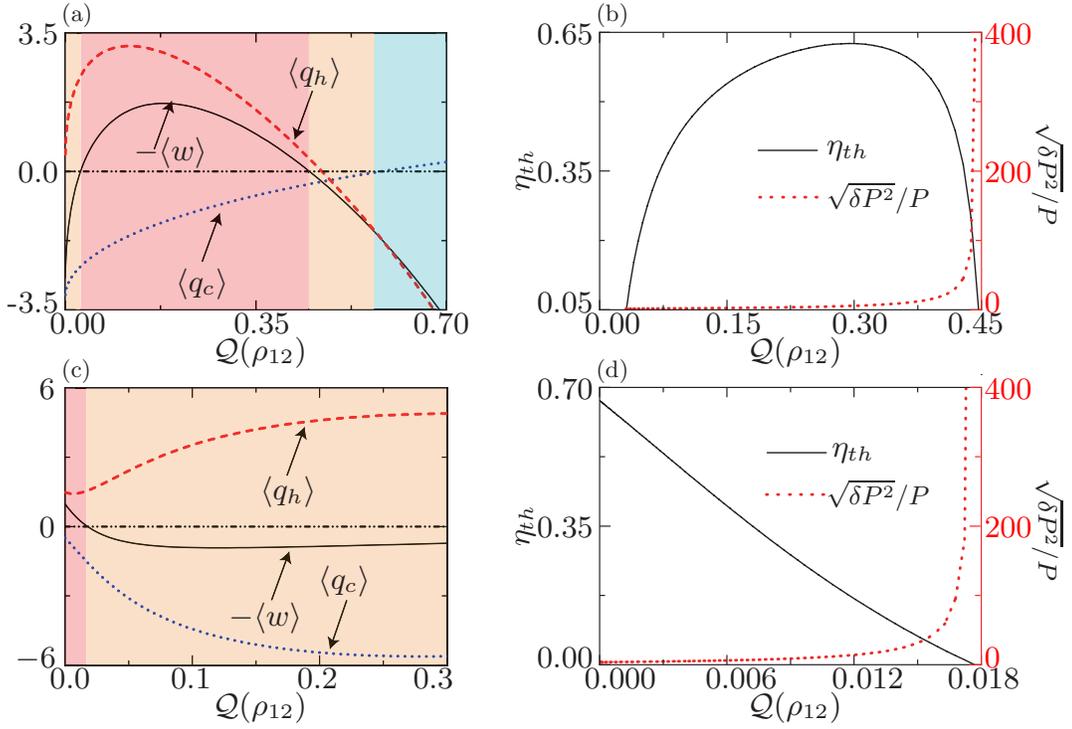
{Here, we consider a quantum Otto engine working with the two-atom system of the Hamiltonian given by Eq. (\ref{ham}) that is driven by two thermal bosonic baths. A machine cycle consisting of two unitary and two isochoric strokes can be realized by modulating the interparticle interactions and the frequency, respectively. Specifically, in the former case, the hot (cold) isochoric stroke proceeds with $\xi=\xi_h$ ($\xi=\xi_c$), and in the latter case, the hot (cold) isochore corresponds to $\omega=\omega_c$ ( $\omega=\omega_h$).}

{When the interaction strength is controlled, the energy
gap is changed during the 
  unitary compression and expansion strokes with the following interaction strength: 
$
    \xi_{\mathrm{com}}=\xi_c(1-t/\tau_{\mathrm{dri}})+\xi_ht/\tau_{\mathrm{dri}},$ and $
     \xi_{\mathrm{exp}}=\xi_h(1-t/\tau_{\mathrm{dri}})+\xi_ct/\tau_{\mathrm{dri}}.
$
The driving Hamiltonian of Eq. (\ref{ham}) becomes $H_{\mathrm{com}}^{sq}(t)=\frac{\omega}{2}(\sigma_1^z+\sigma_2^z)+\xi_{\mathrm{com}}(\sigma_1^+\sigma_2^-+\sigma_1^-\sigma_2^+)$ during the unitary compression and $H_{\mathrm{exp}}^{sq}(t)=\frac{\omega}{2}(\sigma_1^z+\sigma_2^z)+\xi_{\mathrm{exp}}(\sigma_1^+\sigma_2^-+\sigma_1^-\sigma_2^+)$ along the expansion stroke. The system Hamiltonian is thus kept constant at $H^{sq}=H_c^{sq}\equiv \frac{\omega}{2}(\sigma_1^z+\sigma_2^z)+\xi_{c}(\sigma_1^+\sigma_2^-+\sigma_1^-\sigma_2^+)$ during the cold isochore and at $H^{sq}=H_h^{sq}\equiv \frac{\omega}{2}(\sigma_1^z+\sigma_2^z)+\xi_{h}(\sigma_1^+\sigma_2^-+\sigma_1^-\sigma_2^+)$ along the hot isochore. 
The unitary operators in the compression and expansion operations can be written as $U_{ch,hc}=\mathcal{T_>}\{-\frac{i}{\hbar}\int_0^{\tau_\mathrm{dri}} dtH_{\mathrm{com},\mathrm{exp}}^{sq}(t)\}$ with the time-ordering operator $\mathcal{T_>}$. At the end of the hot or cold isochoric stroke, the system reaches the thermal state with $\rho_\alpha^{eq}=e^{-\beta_\alpha H_\alpha^{sq}}/\mathrm{Tr}(e^{-\beta_\alpha H_\alpha^{sq}})$, where $\alpha=c,h$.}

{Because the work is only produced during the two unitary driven strokes, the characteristic function \cite{Fran22} can be given by 
\begin{equation}
\mathcal{X}(u)=\mathcal{X}_{\mathrm{com}}(u)\mathcal{X}_{\mathrm{exp}}(u),
\end{equation}
 where 
$
\mathcal{X}_{\mathrm{com}}(u)=\mathrm{Tr}[e^{iuH_h^{sq}}U_{ch}e^{-\frac{iu}{2}H_c^{sq}}\rho_c^{eq}e^{-\frac{iu}{2}H_h^{sq}}U_{ch}^\dag]$ and $
     \mathcal{X}_{\mathrm{exp}}(u)=\mathrm{Tr}[e^{iuH_c^{sq}}U_{hc}e^{-\frac{iu}{2}H_h^{sq}}\rho_h^{eq}e^{-\frac{iu}{2}H_c^{sq}}U_{hc}^\dag]$. With this function, the average work and work fluctuations can be obtained according to $
-\langle{w}\rangle=i\frac{\partial\mathrm{ln}\mathcal{X}(u)}{\partial u}\big|_{u=0}
$ and $
\delta w^{2}
=-\frac{\partial^2\mathrm{ln}\mathcal{X}(u)}{\partial u^2}\big|_{u=0}
$. The average heat injection during the hot isochoric stroke, which is equivalent to the difference between the system energy at the initial state and that at the final state, is given by 
$ \langle q_h \rangle =\mathrm{Tr}[\rho_h^{eq}H_h^{sq}]-\mathrm{Tr}[U_{ch}\rho_c^{eq}U_{ch}^\dag H_h^{sq}]$. This result, together with the work output $-\langle w\rangle$, determines the efficiency
 $\eta_{th}=-\langle w\rangle/\langle q_h\rangle$.  
In Fig. \ref{qubit}(a), numerical results are presented for the work output ($-\langle w\rangle$) and the heat values absorbed ($\langle q_{c}\rangle$ and $\langle q_h\rangle$) as a function of the quantum discord. The machine operation modes comprise three types: heat engine (red), heater (yellow), and refrigerator (blue). Because of the nonmonotonic behaviors of the work output and the heat injection as a function of the quantum discord, the curve between the efficiency and the discord is parabola-like [c.f. Fig. \ref{qubit}(b)]. The quantum chaos is enlarged by the interactions within the working substance \cite{Hong20}, and thus the relative power fluctuations increase as the quantum discord increases, as also shown in Fig. \ref{qubit}(b).}
 
{The quantum Otto engine can be set up via modulation of its external field (i.e., frequency $\omega$) while keeping $\xi$ constant. During the 
 unitary compression and expansion strokes, the variation of the frequency satisfies the following relationship: $\omega_{\mathrm{com}}=\omega_c(1-t/\tau_{\mathrm{dri}})+\omega_ht/\tau_{\mathrm{dri}},$ and $
     \omega_{\mathrm{exp}}=\omega_h(1-t/\tau_{\mathrm{dri}})+\omega_ct/\tau_{\mathrm{dri}}.$
The driving Hamiltonians of Eq. (\ref{ham}) along the compression and expansion directions are found to be $H_{\mathrm{com}}^{sp}(t)=\frac{\omega_{\mathrm{com}}}{2}(\sigma_1^z+\sigma_2^z)+\xi(\sigma_1^+\sigma_2^-+\sigma_1^-\sigma_2^+)$  and $H_{\mathrm{exp}}^{sq}(t)=\frac{\omega_\mathrm{exp}}{2}(\sigma_1^z+\sigma_2^z)+\xi(\sigma_1^+\sigma_2^-+\sigma_1^-\sigma_2^+)$, respectively. The Hamiltonian along an isochoric remains fixed, and it then reads as $H^{sq}=H_{c}^{sq}\equiv \frac{\omega_c}{2}(\sigma_1^z+\sigma_2^z)+\xi(\sigma_1^+\sigma_2^-+\sigma_1^-\sigma_2^+)$ for the cold isochore and as $H^{sq}=H_h^{sq}\equiv \frac{\omega_h}{2}(\sigma_1^z+\sigma_2^z)+\xi(\sigma_1^+\sigma_2^-+\sigma_1^-\sigma_2^+)$ during the hot isochore. 
By using the same approach that was adopted in the model based on modulation of the interaction strength, we can determine the performance measures, including the work output $-\langle w\rangle$, the efficiency $\eta_{\mathrm{th}}$, and the relative power fluctuations $\sqrt{\delta P^2}/P$, as shown in Figs. \ref{qubit}(c) and \ref{qubit}(d). Given this choice of the parameters, the machine can then only work as a heat engine in the region $\mathcal{Q}(\rho_{12})\le0.018$; otherwise, it operates as a refrigerator. For the heat engine, the quantum discord yields a reduction in efficiency but an increase in  power fluctuations.
We therefore find that when the engines operate with an interacting quantum system, the nonclassical correlations in the working medium affect the machine performance negatively, in contrast to our model, where the quantum correlations in the reservoir lead to improved performance.}

\end{document}